\documentclass{aa}  
\usepackage[dvipsnames]{xcolor}
\usepackage{hyperref}
\hypersetup{
      colorlinks,
      citecolor=blue,   
      linkcolor=blue,
      urlcolor=blue}
\usepackage{subfigure}
\usepackage{graphicx}
\usepackage{booktabs}
\usepackage{tabularx}
\usepackage{tikz}
\usepackage{textcmds}
\usepackage{mathtools}
\usepackage{sidecap}
\usepackage{hyperref}
\usepackage{orcidlink}
\usepackage{makecell}
\usepackage{siunitx}

\usepackage{ulem}
\usepackage{ifthen}

\usepackage[switch]{lineno}

\newcommand{\inlineeq}[4]{%
    \ifthenelse{\equal{#1}{}}{% #1 is empty
        \ifthenelse{\equal{#2}{}}{% #2 is empty
            \ifthenelse{\equal{#3}{}}{% #3 is empty
                \ifthenelse{\equal{#4}{}}{% #4 is empty
                    \errmessage{Error: Nothing is defined}%
                }{% #4 is not empty
                    $\mathrm{#4}$%
                }%
            }{% #3 is not empty 
                \ifthenelse{\equal{#4}{}}{% #4 is empty
                    $#3$%   
                }{% #4 is not empty
                    $#3\, \mathrm{#4}$%
                }%
            }%
        }{% #2 is not empty
            \ifthenelse{\equal{#3}{}}{% #3 is empty
                \ifthenelse{\equal{#4}{}}{% #4 is empty
                    ${#2}$%
                }{% #4 is not empty
                    ${#2}\, \mathrm{#4}$%
                }%
            }{% #3 is not empty 
                \ifthenelse{\equal{#4}{}}{% #4 is empty
                    ${#2}\, #3$%
                }{% #4 is not empty
                    ${#2}\, #3 \, \mathrm{#4}$%
                }%
            }%
        }%
    }{% #1 is not empty
        \ifthenelse{\equal{#2}{}}{% #2 is empty
            \ifthenelse{\equal{#3}{}}{% #3 is empty
                \ifthenelse{\equal{#4}{}}{% #4 is empty
                    \errmessage{Error: no relation defined}% 
                }{% #4 is not empty
                    \errmessage{Error: no relation defined}%
                }%                
            }{% #3 is not empty 
                \ifthenelse{\equal{#4}{}}{% #4 is empty
                    \errmessage{Error: no relation defined}%
                }{% #4 is not empty
                    \errmessage{Error: no relation defined}%
                }%
            }%
        }{% #2 is not empty
            \ifthenelse{\equal{#3}{}}{% #3 is empty
                \ifthenelse{\equal{#4}{}}{% #4 is empty 
                    \errmessage{Error: ill defined relation}%
                }{% #4 is not empty
                    $#1 \, {#2} \, \mathrm{#4}$%
                }%            
            }{% #3 is not empty 
                \ifthenelse{\equal{#4}{}}{% #4 is empty
                    $#1 \, {#2} \, #3$%
                }{% #4 is not empty
                    $#1 \, {#2} \, #3\, \mathrm{#4}$%
                }%
            }%  
        }%
    }%
}%

\newcommand{\msol}{\mathit{M_\odot}}
\newcommand{\rsol}{\mathit{R_\odot}}

\newcommand{\RNum}[1]{\uppercase\expandafter{\romannumeral #1\relax}}

\begin{document} 
   \title{From spherical stars to disk-like structures: 3D common-envelope evolution of massive binaries beyond inspiral}
   \author{Marco Vetter\inst{1}\fnmsep\inst{3}\orcidlink{0009-0007-2322-6001}
          \and
          Friedrich K. R{\"o}pke\inst{1}\fnmsep\inst{2}\fnmsep\inst{3}\orcidlink{0000-0002-4460-0097}
          \and 
          Fabian R. N. Schneider\inst{1}\fnmsep\inst{3}\orcidlink{0000-0002-5965-1022}
          \and
          R{\"u}diger Pakmor\inst{4}\orcidlink{0000-0003-3308-2420}
          \and
          Sebastian~T.~Ohlmann\inst{5}\orcidlink{0000-0002-6999-1725}
          \and
          Mike Y. M. Lau\inst{3}\orcidlink{0000-0002-6592-2036}
          \and
          Robert Andrassy\inst{1,3}
          }

   \institute{Zentrum f{\"u}r Astronomie der Universit{\"a}t Heidelberg, Astronomisches Rechen-Institut, M{\"o}nchhofstr, 12--14, 69120 Heidelberg, Germany\\
              \email{marco.vetter@stud.uni-heidelberg.de}
        \and
             Zentrum f{\"u}r Astronomie der Universit{\"a}t Heidelberg, Institut f{\"u}r Theoretische Astrophysik, Philosophenweg 12, 69120 Heidelberg, Germany\\
        \and
            Heidelberger Institut f{\"u}r Theoretische Studien, Schloss-Wolfsbrunnenweg 35, 69118 Heidelberg, Germany\\
        \and
            Max-Planck-Institut f{\"u}r Astrophysik, Karl-Schwarzschild-Str. 1, D-85748, Garching, Germany\\
        \and
            Max Planck Computing and Data Facility, Gießenbachstraße 2, 85748 Garching, Germany\\
    }
        
   \date{Received YYY; accepted ZZZ}

  \abstract
  {Self-consistent three-dimensional modeling of the entire common-envelope phase of gravitational wave progenitor systems until full envelope ejection is challenged by the vast range of spatial and temporal scales involved in the problem. Previous attempts were either terminated shortly after the rapid spiral-in with significant amounts of gravitationally bound material left in the system or they omitted this plunge-in phase and modeled the system afterward. We investigated the common-envelope interactions of a $10 \, \msol$ red supergiant primary star with a black hole and a neutron star companion, respectively, until full envelope ejection ($\, {\gtrsim}\, 97 \, \mathrm{\%}$ of the envelope mass). 
  In contrast to the expectation from e.g.~population synthesis models, we find that the dynamical plunge-in of the systems determines (to leading order) the orbital separations of the core binary system, while the envelope ejection by recombination acts only at later stages of the evolution and fails to harden the core binaries down to orbital frequencies where they qualify as progenitors of gravitational-wave-emitting double-compact object mergers.
  Diverging from the conventional picture of an expanding common envelope that is ejected more or less spherically, our simulations show a new mechanism: The rapid plunge-in of the companion transforms the spherical morphology of the giant primary star into a disk-like structure. During this process, magnetic fields are amplified, and the subsequent transport of material through the disk around the core binary system drives a fast jet-like outflow in the polar directions. While most of the envelope material is lost through a recombination-driven wind from the outer edge of the disk, about $7\, \mathrm{\%}$ of the envelope leaves the system via the magnetically driven outflows.
  We further explored the potential evolutionary pathways of the post-common-envelope systems in light of the expected remaining lifetime of the primary core ($2.97\, \msol$) until core collapse ($6{\times}10^{4}\, \mathrm{yr}$), most likely forming a neutron star. We find that the interaction of the core binary system with the circumbinary disk substantially increases the likelihood of giving rise to a double-neutron star merger ($55\, \mathrm{\%}$) or a neutron star black hole ($5\, \mathrm{\%}$) merger event.}


   \maketitle
%

%\linenumbers
%%%%%%%%%%%%%%%%%%%%%%%%%%%%%%%%%%%%%%%%%%%%%%%%%%%%%%%%%%%%%%%%%%%%%
\section{Introduction}\label{sec:intro}
Our understanding of the progenitor systems leading to gravitational-wave (GW) emitting mergers of double compact objects is challenged by several major knowledge gaps \citep[e.g.,][]{dominik2012a, ivanova2013a, belczynski2020a,tauris2017a, vigna2018a,giacobbo2018a}. Two of which are the reaction of the system to the supernova (SN) explosion forming the compact objects and the ability of a common-envelope (CE) phase to shrink the orbit sufficiently for GW emission, initiating a merger within a Hubble time.

In the CE interaction, the more compact secondary object is engulfed by the envelope of the giant primary star due to a preceding dynamically unstable mass transfer \citep[MT, e.g.,][]{eggleton2011a} phase or a tidal interaction such as the Darwin instability \citep{darwin1879a}, for example.
When orbiting inside the CE, the companion and the core of the primary star experience drag forces caused by the gravitational pull of accumulated envelope material behind them through self-gravity \citep{chandrasekhar1943a,dokuchaev1964a,ostriker1999a,kim2007a,kim2010a}. These drag forces tighten the orbit of the binary system that formed from the core of the primary star and the companion (referred to as \qq{core binary system} in the following). During this process, the core binary system  transfers angular momentum and orbital energy to the envelope material, which eventually leads to a partial or full ejection of the CE.

The motion of the companion object and the core of the primary star within the envelope material triggers numerous (magneto-)hydrodynamical instabilities  that not only affect the final orbital separation, but also shape the flows of the magnetized plasma. The implied dynamics create unique features such as magnetic outflows perpendicular to the orbital plane \citep[][]{ondratschek2022a} and the potential formation of a centrifugally supported structure interacting with the inner post-CE binary during and after the dynamic CE interaction \citep{roepke2022a,gagnier2023a,tuna2023a,wei2023a}.

Given the vast range of spatial and temporal scales involved in the problem -- spanned by the scales of the progenitor star and the compact companion -- three-dimensional (3D) simulations of massive CE systems face significant difficulties in following the CE phase to full envelope ejection \citep[e.g.,][]{law-smith2020a, moreno2022a, lau2022a, lau2022b}. During the simulated time span, these models fail to tighten the central binary to distances at which the emission of GWs becomes important, as predicted by the analytical prescription \citep[e.g.,][]{webbink1984a} used in population synthesis models \citep[e.g.,][]{hurley2002a,dominik2012a,stevenson2017a,vigna2018a,mapelli2018a,kruckow2018a,belczynski2020a}. 
This strong discrepancy between theoretical expectations and numerical findings has to be settled in order to make reliable rate predictions of GW-induced mergers of compact objects. Following the discussion of \citet{moreno2022a}, their simulations appear to be converged in the final orbital separations. In this case, two causes for the discrepancy are conceivable: First, the current simulations fail to capture all relevant physical effects in the plunge-in phase of the CE interaction, and second, there are mechanisms after the rapid spiral-in sufficiently contracting the core binary until the system reaches full envelope ejection. 

Here, we investigate the second case. We report on the extensions of the simulations conducted by \cite{moreno2022a} of two CE interactions of a \inlineeq{}{}{10}{\msol} post core-helium burning red super-giant (RSG) primary star (at an age of \inlineeq{}{}{25.5}{\mathrm{Myr}}) with a black hole  (BH) and a neutron star (NS) companion. After a SN explosion of the core of the primary star \citep[most likely forming a NS,][]{podsiadlowski2004a,schneider2021a}, the remaining compact binary system would be a promising progenitor for a NS-BH or NS-NS merger event \citep[e.g.,][]{podsiadlowski2004a,vigna2018a,giacobbo2018a,belczynski2020a,mandel2021a}. 
However, \citet{moreno2022a} found a final orbital separation of \inlineeq{a_\mathrm{f,\, NS}}{=}{16}{\rsol} for the NS companion and \inlineeq{a_\mathrm{f,\, BH}}{=}{47}{\rsol} for the BH companion at the end of their simulation, which are both too wide for the systems to merge within a Hubble time by GW emission. 
They further showed that even taking into account a subsequent second MT phase as well as an isotropically distributed natal NS kick, only \inlineeq{}{}{8.7}{\mathrm{\%}} of the NS companion systems and \inlineeq{}{}{0.6}{\mathrm{\%}} of the BH companion systems are expected to form a NS-NS or NS-BH merger, respectively. We mitigated the tremendous computational cost encountered in the simulations conducted by \citet{moreno2022a} with a new refinement approach of gas cells deep in the gravitational potential of the primary core and companion.

We summarize our model and methods in Sect.~\ref{sec:methods} and show therein the pivotal aspects of our initial primary star (Sect.~\ref{sec:initial_setup}), CE model (Sect.~\ref{sec:ce-phase}), and introduce the new refinement approach (Sect.~\ref{sec:refinement}). The results of our work are shown in Sect.~\ref{sec:results}, where we describe the temporal evolution of the orbital separations and the envelope ejection (Sect.~\ref{sec:oe}), the appearance of a geometrically thick toroidal structure and magnetically driven bipolar outflows (Sect.~\ref{sec:morpho}, where the latter is described in more detail in the upcoming publication Vetter et al.~in prep.), the identification of the torus as a circumbinary disk (CBD) and its characteristic properties (Sect.~\ref{sec:CBD}), updated $\alpha_\mathrm{CE}$ values (Sect.~\ref{sec:results_alpha_CE}), and possible final fates of the systems (Sect.~\ref{sec:final_fate}) based on the outcome of our simulations. Last, we discuss our results in Sect.~\ref{sec:discussion} and conclude in Sect.~\ref{sec:conclusion}.    
 
%%%%%%%%%%%%%%%%%%%%%%%%%%%%%%%%%%%%%%%%%%%%%%%%%%%%%%%%%%%%%%%%%%%%%
\section{Methods}\label{sec:methods}
We use the identical CE models as presented by \citet{moreno2022a}, where they closely adhere to the procedures outlined in 
\citet{ohlmann2016a}, \citet{ohlmann2017a}, and \citet{sand2020a} for conducting CE simulations. We refer to these publications for a more detailed discussion and additional information, and only summarize basic aspects of both the initial primary star and the CE models in Sect.~\ref{sec:initial_setup} and Sect.~\ref{sec:ce-phase}. 

The three-dimensional simulations shown in this work are conducted with the 3D moving-mesh magnetohydrodynamics code \textsc{arepo} \citep{springel2010a, pakmor2011d, pakmor2013b}, which employs a second-order finite-volume method. The Powell scheme \citep{powell1999a, pakmor2013b} is utilized to control the divergence of the magnetic field. Newtonian self-gravity was calculated with a tree-based algorithm. To account for recombination energy, we applied the OPAL \citep{iglesias1996a, rogers1996a, rogers2002a} equation of state (EoS) similar to previous CE simulations conducted by, for example, \citet[][]{sand2020a, kramer2020a}, and \citet{moreno2022a}.

Here, we introduce a new refinement approach (Sect.~\ref{sec:refinement}) for the cells within the softened potential around the primary core and companion \citep[][]{ohlmann2016a} to ease the computational costs encountered in the late stages of the simulations presented by \citet{moreno2022a}, which prevented a continuation to later times in the evolution of the considered systems.

%--------------------------------------------------------------------
\subsection{Establishing a stable primary star model}\label{sec:initial_setup}
For our initial primary star model, we consider a \inlineeq{M_1}{=}{10}{\msol} zero age main sequence star with solar metallicity (\inlineeq{Z}{=}{0.02}{}). The star is first evolved with the one dimensional \textsc{mesa} code \citep{paxton2011a,paxton2013a,paxton2015a,paxton2018a,paxton2019a} to a RSG with an age of about \inlineeq{}{}{25.5}{\mathrm{Myr}}. At the desired evolutionary stage, the star has exhausted its core-helium abundance and already developed a deep convective envelope \citep[][Fig.~1]{moreno2022a}. The considered stellar model has a mass of \inlineeq{M_1}{=}{9.4}{\msol}, a radius of \inlineeq{R_1}{=}{395}{\rsol}, an effective temperature of \inlineeq{T_\mathrm{eff}}{=}{3564}{\mathrm{K}} and a logarithmic luminosity of \inlineeq{\log{L/L_\mathrm{\odot}}}{=}{4.35}{}.

Subsequently, the one-dimensional stellar profile is mapped onto the 3D grid of \textsc{arepo} closely following the procedure outlined in \citet{ohlmann2016a}. During this process, two main challenges arise: First, we have to represent the inner core region, which is very small compared with the overall spatial scales of the RSG. As discussed by \citet{moreno2022a}, resolving this core also implies a severe timescale problem. Second, the unavoidable introduction of discretization errors perturbs the star from its hydrostatic equilibrium (HSE) and gives rise to spurious velocities, which have to be kept under control.

The first challenge is addressed by replacing the star's core with a point mass that interacts with the other material only via gravity \citep[Section 3.2]{ohlmann2017a}. In this context, the core of the star is defined based on the cut radius $R_\mathrm{cut}$, which is ideally chosen such that (i) most of the convective envelope is resolved on the \textsc{arepo} grid (the bottom of the convective envelope is at \inlineeq{r}{=}{17.9}{\rsol} with mass coordinate \inlineeq{m}{=}{3.07}{\msol}), (ii) it encompasses key points, such as the compression point 
(\citealp{ivanova2011a}; \inlineeq{r}{=}{0.40}{\rsol}, \inlineeq{m}{=}{2.78}{\msol}) and the location where the hydrogen mass fraction X drops below $0.1$ 
\citep{dewi2000a} (\inlineeq{r}{=}{0.33}{\rsol}, \inlineeq{m}{=}{2.76}{\msol}) and ideally (iii) below or near the final orbital separation to ensure convergence \citep{ohlmann2017a,moreno2022a}. The points in (ii) collectively define the amount of envelope material which needs to be ejected to prevent immediate re-expansion of the star. 

While a cut radius on the order of criterion (ii) is beyond the capabilities of current numerical approaches, conditions (i) and (iii) can be met, and \citet{moreno2022a} find a hydrostatically stable primary star and converging results in the final orbital separations in their relaxation and CE simulations, respectively. The considered primary model was set up with \inlineeq{R_\mathrm{cut}}{\approx}{20}{\rsol} and a (linear) resolution of the softened sphere of \inlineeq{N_\mathrm{cps}}{=}{40}{} grid cells (\qq{cps} stand for \qq{cells per softening length,} i.e., the actual spatial resolution of the softened volume is approximately given by $N_\mathrm{cps}^{3}$). With this choice, \inlineeq{}{}{97}{\mathrm{\%}} of the mass of the hydrogen-rich envelope is contained on the hydrodynamic grid. The gas cells within the core region are adapted to the solution of a modified Lane-Emden equation to ensure a smooth transition of the gradients over the cut radius \citep{ohlmann2017a}.

The simulated 3D domain extends far beyond the radius of the star and spans \inlineeq{}{}{3159}{\rsol} in each spatial dimension. The star is embedded in a \qq{pseudo-vacuum} with a density of \inlineeq{}{}{10^{-15}}{\mathrm{g\, cm^{-3}}} and a temperature of \inlineeq{}{}{4000}{\mathrm{K}}. We utilize a Lagrangian refinement criterion for the gas cells outside the softened spheres of point particles, that is, the mass of all cells (on average, our simulation contains $4,569,947$ grid cells) fluctuates by less than a factor of two around a target mass resolution of \inlineeq{m_\mathrm{cell}}{=}{1.6{\times}10^{-6}}{\msol}. It is important to note that during the CE simulation, the simulated box expands far beyond the initially set box size, and the number of cells contained in our domain also varies given the applied mass refinement criterion. 

\subsection{Common-envelope phase}\label{sec:ce-phase}
We aim to simulate two CE interactions with secondary masses of \inlineeq{M_2}{=}{5}{\msol} and \inlineeq{M_2}{=}{1.4}{\msol}, implemented as gravity-only point particles (similar to the core of the primary star). The companion masses are chosen to represent a BH and a NS companion, respectively. With the implementation of the companions as point particles, we neglect potential accretion, neutrino cooling and jet formation through accretion streams onto the compact objects in our simulations.  
The companions are initially in full corotation with the envelope material.\footnote{While this condition would physically prevent the onset of the CE phase, simulations conducted by \citet{moreno2022a} demonstrated a successful loss of corotation, given that the primary star has not been relaxed in the combined potential of the binary. In line with our objective of replicating and extending these simulations, we also maintain this initial condition in our models.} 

A seed dipolar magnetic field with a surface field strength of \inlineeq{}{}{10^{-6}}{\mathrm{G}} is applied to the primary star in all conducted simulations \citep{ohlmann2016b, moreno2022a}.
To quantify the amount of ejected (i.e., unbound) envelope material during the CE event, we adhere to the commonly used energy criteria \citep[e.g.,][]{ohlmann2016a, sand2020a, moreno2022a,lau2022a,lau2022b,chamandy2023a}, which read, 
\begin{align}
    e_\mathrm{kin} + e_\mathrm{pot} > 0 , \quad &\textrm{kinetic-energy criterion}, \label{eq:e_kin_crit}\\
    e_\mathrm{kin} + e_\mathrm{therm} + e_\mathrm{pot} > 0 , \quad &\textrm{thermal-energy criterion}, \label{eq:e_therm_crit}\\
    e_\mathrm{kin}+ e_\mathrm{int} + e_\mathrm{pot} > 0 , \quad &\textrm{internal-energy criterion},\label{eq:e_int_crit}
\end{align}
where $e_\mathrm{kin}$ is the kinetic, $e_\mathrm{therm}$ is the thermal, $e_\mathrm{int}$ is the internal energy (including thermal and ionization energy), and $e_\mathrm{pot}$ is the potential energy, respectively. We express the amount of unbound material by the sum over all cell masses, which satisfy the individual criteria, and relative to the total envelope mass \inlineeq{f_\mathrm{ej}}{=}{M_\mathrm{unbound}/M_\mathrm{env}}{} (here, $M_\mathrm{env} \, {\approx} \,  6.648 \, M_\odot$ is defined by the total gas mass contained on the \textsc{arepo} grid in our simulations).

In contrast to \citet{moreno2022a}, we additionally employ Lagrangian tracer particles in our simulations to post-process the gas flows, since we are still restricted by the finite-volume approach of \textsc{arepo}. These particles are implemented in \textsc{arepo} as passively advected Lagrangian particles. The trajectories are advected with the local velocity field interpolated to the position of the tracer particles \citep{genel2013a}. In all simulations, we add a total of \inlineeq{N_\mathrm{tr}}{=}{799,212}{} tracer particles (i.e., approximately a sixth of the initial number of hydro-cells) that sample the initial mass distribution over all hydro-cells in the entire simulation box. Given the target mass resolution in our simulations, each tracer particle represents the same mass of \inlineeq{m_\mathrm{tr}}{=}{M_\mathrm{env}/N_\mathrm{tr} \,{\approx}\, 8.32 {\times} 10^{-6}}{\msol}.

\begin{figure*}
    \centering
    \resizebox{\hsize}{!}{
    \includegraphics{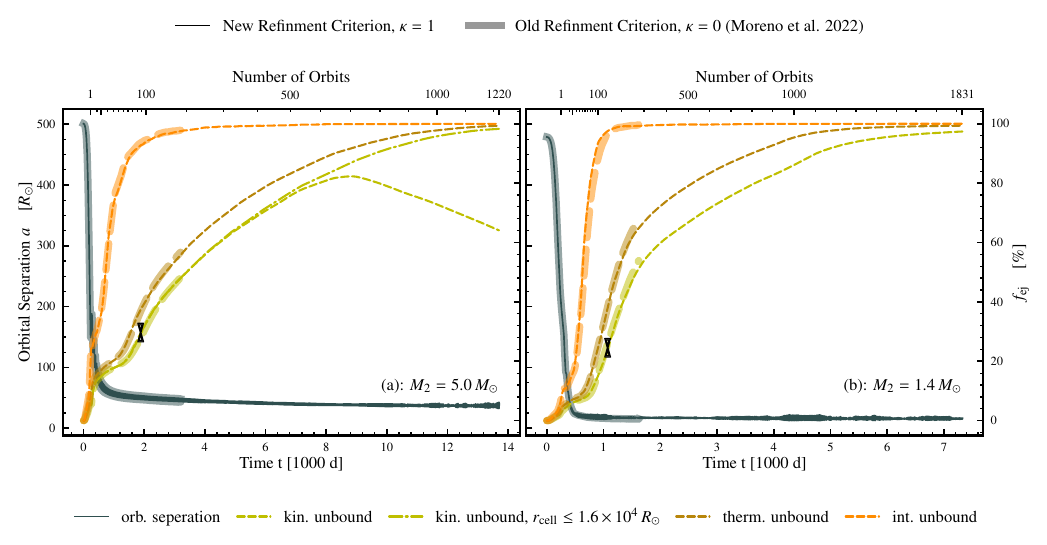}}
    \caption{Temporal evolution of the orbital separation $a$ (solid dark) and the fraction of unbound material $f_\mathrm{ej}$ according to the kinetic (dashed green, Eq.~\ref{eq:e_kin_crit}), thermal (dashed rust, Eq.~\ref{eq:e_therm_crit}), and internal energy (orange, Eq.~\ref{eq:e_int_crit}) criterion, respectively. The simulation involving the BH is shown on the left, while the NS scenario is on the right. The thin lines in the plots represent the simulations that contained the new refinement approach (see Sect.~\ref{sec:refinement}, $\kappa \,{=}\, 1$), while the thick, faint lines correspond to the results of \citet{moreno2022a}. On the left-hand side, we additionally show the unbound mass fraction only for cells contained within a radius of $R\,{=}\,1.6{\times} 10^4 \, \rsol$ around the central binary due to the artificial \qq{rebinding} (see Sect.~\ref{sec:oe}) of material in the pseudo-vacuum for $t \, {\ge }\, 4400 \, \mathrm{d}$ (see Sect.~\ref{sec:oe}). Furthermore, the hourglass markers represent the individual starting points of the magnetically driven outflow in the simulations with $\kappa \,{=}\, 1$. The minor ticks in both plots in between $1\, \textrm{--} \, 10$ orbits are adapted for readability, and only the 5 orbit mark is shown.} 
    \label{fig:dyn_evolution}
\end{figure*}
%--------------------------------------------------------------------
\subsection{New refinement criterion for cells inside the softening length of the point particles}\label{sec:refinement}
As already briefly mentioned in Sect.~\ref{sec:initial_setup}, the primary-core approximation \citep{ohlmann2016a} requires a sufficient number of cells per softening length $N_\mathrm{CPS}$ for the star to establish HSE in the relaxation run. However, during the spiral-in of the companion, the softening length $h$ is reduced proportional to the orbital separation $a$ between the point particles as soon as the separation decreases to \inlineeq{a(t)}{=}{2.5h(t)}{} (enforcing \inlineeq{h(t)}{\le}{0.4a(t)}{} for any given time)\footnote{The gravitational interaction between the point particles is found to be resolved sufficiently when the combined softening lengths are at most a fifth of the current orbital distance \citep[][]{ohlmann2016b}.}. This behavior ensures that the softened spheres of the core and secondary do not overlap.
In previous works \citep[e.g.,][]{ohlmann2016a,ohlmann2016b, ohlmann2017a, sand2020a, moreno2022a}, the  number of cells per softening length was kept constant during the CE interaction. This decreases the cell sizes inside the softened gravitational potential, leading to an increased physical resolution within the approximated core region. According to the Courant--Friedrichs--Lewy (CFL) condition \citep{courant1928a}, the gradual decline in size of the cells reduces the time step in the binary simulation.

This problem becomes even more evident in the final stages of the CE interaction, when the core is in a close orbit with the companion and detached from its former envelope material. The necessity of the forced decrease in cell size caused by the constant resolution \inlineeq{N_\mathrm{CPS}}{=}{\mathrm{const}}{} criterion seems to be questionable, since we neither have to ensure HSE anymore nor are we able to resolve the core in the first place.
As a result of the decreased cell sizes, the simulations of \citet{moreno2022a} became prohibitively expensive after the spiral-in due to too small time steps and only \inlineeq{f_\mathrm{ej, \, BH}}{\approx}{48}{\mathrm{\%}} (\inlineeq{f_\mathrm{ej,\, NS}}{\approx}{54}{\mathrm{\%}}) of the envelope mass in the CE simulation with the BH (NS) companion was unbound according to the kinetic energy criterion (Eq.~\ref{eq:e_kin_crit}) over the affordable simulation time.  

To ease the computational costs in our simulations, we thus follow the idea of controlling the number of cells per softening length $N_\mathrm{CPS}$ by establishing a new refinement criterion, for which the number of cells per softening length is a power law of $h$, 
\begin{align}
	N_\mathrm{CPS} \propto h^\kappa \quad \Rightarrow \quad N_\mathrm{CPS}(t) = N_\mathrm{CPS,\, 0} \left (\frac{h(t)}{h_0}\right )^\kappa,\label{eq:N_CPS}
\end{align}
where $N_\mathrm{CPS,\,0}$ is the number of initially used cells per softening length and $h_0$ the initial softening length, which we set to the cut radius $R_\mathrm{cut}$ of our relaxed primary star model in Sect.~\ref{sec:initial_setup}. 
We note that \inlineeq{\kappa}{=}{0}{} represents the standard refinement criterion with \inlineeq{N_\mathrm{CPS}}{=}{N_\mathrm{CPS,\, 0} \, {=}\, \mathrm{const}}{} used in previous works, while we use \inlineeq{\kappa}{=}{1}{} in this work.

Ultimately, \inlineeq{\kappa}{=}{1}{} leads to an approximately constant cell size inside the softening length throughout the CE simulation and thus increases the lowest time step compared to \inlineeq{\kappa}{=}{0}{} (see Fig.~\ref{fig:appendixB_comp_costs} and Appendix~\ref{sec:appendix_numerical_uncertanties} in general for more information). Neither \inlineeq{\kappa}{=}{0}{} nor \inlineeq{\kappa}{=}{1}{} are ideal solutions and are born out of the necessity to approximate the primary core and companion and in future studies, a more sophisticated model is desirable. 

\begin{figure}
    \centering
    \resizebox{\hsize}{!}   {\includegraphics{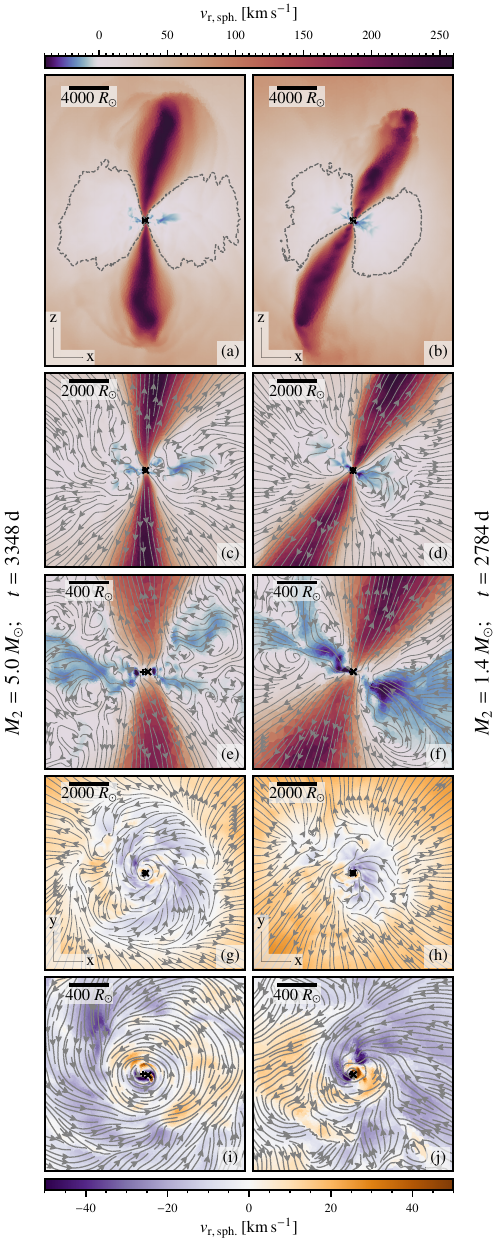}}
    \caption{Slices of spherical radial velocity $v_\mathrm{r,\, sph.}$ in the $x\, \textrm{--} \, z$ plane in (a)-(f) and in the $x \, \textrm{--}\, y$ plane in (g)-(j) for the BH (NS) companion in the left (right) column. The iso-surfaces in (a) and (b) enclose the bound mass according to the kin.~energy criterion (Eq.~\ref{eq:e_kin_crit}) in dashed gray. The gray streamlines in (c)-(j) represent the velocity field in the corresponding plane. The primary core (companion) is marked with a black cross (plus sign).}
    \label{fig:disk_summary}
\end{figure}

%%%%%%%%%%%%%%%%%%%%%%%%%%%%%%%%%%%%%%%%%%%%%%%%%%%%%%%%%%%%%%%%%%%%%
\section{Results}\label{sec:results}
 
%--------------------------------------------------------------------
\subsection{Global orbital evolution and envelope ejection}\label{sec:oe}
Once the companions are placed above the surface of the primary star with initial separations of $a_\mathrm{i,\, BH} \, {=}\, 501 \, \rsol$ and $a_\mathrm{i,\, NS}\,{=}\, 479 \, \rsol$ for the BH and NS companion, respectively, we observe a similar dynamic behavior during the plunge-in and rapid spiral-in phases as \citet{moreno2022a}.
Consequently, we find a good overall agreement between the simulations with $\kappa\,{=}\,0$ and the simulations with the new criterion with $\kappa\,{=}\,1$ in the orbital evolution (almost perfect correlation with only small deviations, see Appendix~\ref{sec:appendix_numerical_uncertanties}, Fig.~\ref{fig:compare_sims_bland_altman}) as well as in the fraction of ejected envelope material $f_\mathrm{ej}$ as shown in Fig.~\ref{fig:dyn_evolution}.
However, in the late-time evolution, both systems (especially the NS scenario) develop small changes in eccentricity (and even orbital precession in the system involving a NS companion, that is, Fig.~\ref{fig:disk_summary}b). At present, there is no clear explanation for this behavior, and we refer to Sect.~\ref{sec:discussion_oe} and Appendix~\ref{sec:appendix_numerical_uncertanties} for further discussion.

We were able to follow the dynamical CE phase to almost full envelope ejection (more than $97\, \mathrm{\%}$ of the former envelope in both systems is formally ejected according to Eq.~\ref{eq:e_kin_crit}) which was established after \inlineeq{}{\approx}{$\num{13708}$}{\mathrm{d}} for the BH companion and \inlineeq{}{\approx}{$\num{7330}$}{\mathrm{d}} for the NS companion, respectively (Fig.~\ref{fig:dyn_evolution}).
In the BH companion case, we observe an increasing portion of material that becomes bound again in the pseudo-vacuum (in the following referred to as \qq{rebinding of material}) at $R\,{>}\,3{\times} 10^5 \, \rsol$.
There, the already formally ejected envelope material is forced to dissipate its kinetic energy to the ambient low-density material and, hence, appears to be bound again according to the kinetic energy criterion (Fig.~\ref{fig:dyn_evolution}a, dashed green line). Notably, the immediate fallback of this material is prevented by a pressure support established through the later on ejected material. A similar observation has also been made in the simulations conducted by \citet{chamandy2019a,chamandy2023a}. Here, we consider it an artifact caused by the need to fill the outer regions of our simulation domain with low-density pseudo-vacuum. 
We correct for this unexpected rebinding effect in the fraction of unbound envelope material $f_\mathrm{ej}$  by considering cells with distances larger or equal to $1.6{\times}10^{4}\, \rsol$ with respect to the central core binary system as unbound independently of the result Eq.~\ref{eq:e_kin_crit} (Fig.~\ref{fig:dyn_evolution}a, dash-dotted green line). 

Upon almost full envelope ejection, the conducted CE simulations result in a final orbital separation of $a_\mathrm{f, \, BH}\,{\approx}\, 37 \, \rsol$ (after 1220 orbits) for the BH scenario, and  $a_\mathrm{f,\, NS}\,{\approx}\, 14 \,  \rsol$ for the NS companion establishes (after 1831 orbits).
Compared with the final distances reported by \citet{moreno2022a} of $a_\mathrm{f, \, BH}\,{\approx}\, 47 \, \rsol$ and $a_\mathrm{f,\, NS}\,{\approx}\, 15 \, \rsol$, the core binary hardened by \inlineeq{}{\sim}{21}{\%} in the BH companion case and its distance changes only marginally for the CE interaction involving the NS companion.

%--------------------------------------------------------------------
\subsection{Reshaping of the envelope} 
\label{sec:morpho}
Similar to the low-mass system studied by \citet{ondratschek2022a}, the seed magnetic field rapidly amplifies during the pre-plunge and in-spiral phases of the CE interaction \citep[][]{ohlmann2016b} and leads to the formation of a magnetically driven bipolar outflow (see Fig.~\ref{fig:dyn_evolution}, black marker) perpendicular to the orbital $x\, \textrm{--} \, y$ plane. This feature is best visible in the spherical radial velocity $v_\mathrm{r, \, sph.}$ (the subscript \qq{sph} stands for the calculation in spherical coordinates) in Fig.~\ref{fig:disk_summary}a and \ref{fig:disk_summary}b for the BH and NS companion, respectively. The accelerated material reaches peak radial velocities of almost ${\sim}\, 300\, \mathrm{km\, s^{-1}}$.
Moreover, the rapidly expelled material forms characteristic bow shocks (Fig.~\ref{fig:disk_summary}a and \ref{fig:disk_summary}b) resulting from the interaction with the already ejected envelope \citep[similar to the findings of][]{ondratschek2022a}.\footnote{We analyze the magnetically driven outflow and its origin in detail in a follow-up publication (Vetter et al.~in prep.).}

Prior to the onset of the bipolar outflow, the bound material in the simulations (approximately $ 68\, \mathrm{\%}$ [BH companion] and $ 76\, \mathrm{\%}$ [NS companion] of the total envelope mass according to Eq.~\ref{eq:e_kin_crit}) redistributes to a thick toroidal structure in the orbital plane of the core binary (see gray dashed contour in Fig.~\ref{fig:disk_summary}a and b) and together with the bipolar outflows show morphological similarities to bipolar pre-planetary nebulae (PNe) observed in lower-mass systems.

Furthermore, radially in-falling regions onto the central binary develop (see blue areas throughout the entire Fig.~\ref{fig:disk_summary}), suggesting the formation of a thick circumbinary accretion disk (Sect.~\ref{sec:CBD}) refueling the magnetically driven outflows as already proposed but not confirmed in \citet{ondratschek2022a}. The velocity fields in the orbital plane exhibit circularized motion (Fig.~\ref{fig:disk_summary}i and j) and even show indications for the presence of turbulence (Fig.~\ref{fig:disk_summary}e and f), possibly enabling angular momentum and mass transport along the orbital $x \, \textrm{--} \, y$ plane.

The circular motion of the bound material transitions into a radially outward pointing motion of the gas for radii larger than $R\,{\approx}\, 1000\, \textit{--}\,  4000\, \rsol$ with respect to the core binary system (e.g., in Fig.~\ref{fig:disk_summary}g and \ref{fig:disk_summary}h, the transition is around $R\,{\approx}\, 2000\, \rsol$ for the BH companion and $R\,{\approx}\, 1000\, \rsol$ for the NS companion).
This transition region appears to coincide with that of hydrogen recombination (see Movie M2 and Movie M4 in Table~\ref{tab:movietable} in Appendix~\ref{sec:appendix_movies}) and we find that the ejection of this material is mainly driven by heating from recombination of helium and hydrogen in the cool outer parts of the disk (as we subsequently explain in more detail in Sect.~\ref{sec:CBD}) and resembles a radial wind with radial velocities of \inlineeq{}{\approx}{40\,\textit{--}\,50}{\mathrm{km \, s^{-1}}}. In fact, the envelope ejection at that stage is completely dominated by the radially outflowing material, with progressively decreasing mass loss rates on the order of $\dot{M}\,{\propto}\, 10^{-4}\,\msol\, \mathrm{d^{-1}}$ down to $10^{-5}\, \msol\, \mathrm{d^{-1}}$ as the simulations evolve toward full envelope ejection. The point in time where the radial winds become dominant in the CE ejection is visible by the steeply increasing fraction of ejected material $f_\mathrm{ej}$ according to Eq.~(\ref{eq:e_kin_crit}) in Fig.~\ref{fig:dyn_evolution}a (\ref{fig:dyn_evolution}b) around ${\sim}\, 1400 \, \mathrm{d}$ (${\sim}\, 700 \, \mathrm{d}$) for the BH (NS) case. The reshaped envelope is sufficiently adiabatically cooled at this point that recombination commences. 

In summary, we observe a morphological transformation of the envelope due to orbital energy and angular momentum transferred during the spiral-in. The initial spherical primary star is reshaped into a toroidal structure surrounding the core binary (for example, see Movie~M1-M4 in Table~\ref{tab:movietable}). The foundation for a CBD might thus have already been set as early as the end of the plunge-in phase (i.e., within the first \inlineeq{}{\approx}{800}{\mathrm{d}} for the BH and \inlineeq{}{\approx}{500}{\mathrm{d}} for the NS companion) and we further describe this in Sect.~\ref{sec:CBD}. 
The CBD is affected by two primary mass-loss channels. First, the inward pointing mass flux indicated by the negative radial velocities of the gas and second, the radial winds driven by recombination. The first channel is responsible for refueling the magnetically propelled jet-like outflows perpendicular to the orbital plane, which, together with the bound torus, resemble the morphological structure of bipolar pre-PN.

%--------------------------------------------------------------------
\subsection{Formation of a circumbinary disk and recombination winds}\label{sec:CBD}
\begin{figure*}
    \includegraphics[width=\linewidth]{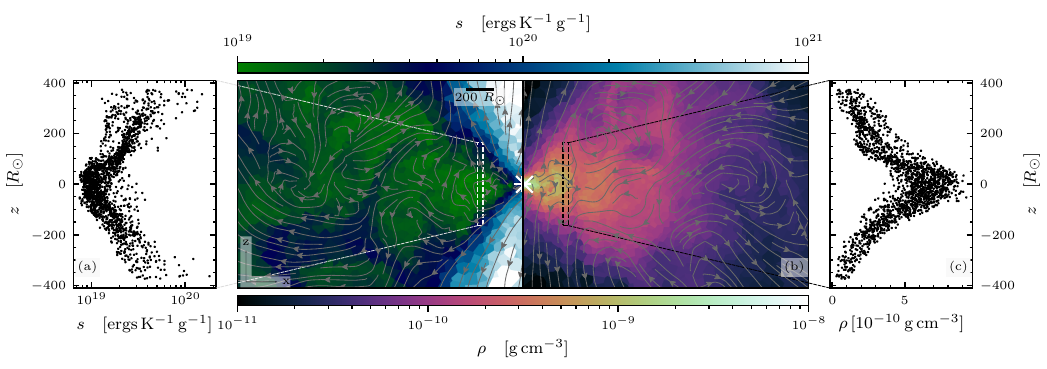}
    \caption{Specific entropy $s$ and density $\rho$ slices of the simulation with the BH companion at $t\,{=}\, 8244\, \mathrm{d}$. The quantities are shown to the left ($x\leq 0$ for the spec. entropy) and right ($x \geq 0$ for the density) in an edge-on view ($x\,\textit{--}\, z$) in (b), respectively. Furthermore, the distribution of all cells along the $z$-direction in a cylindrical shell (indicated by the dashed rectangle) in (a) and (c). The gray streamlines in (b) show the velocity field in the $x\,\textit{--}\, z$ plane.}
    \label{fig:overview_disk}
\end{figure*}
During the rapid in-spiral of the companion, the envelope of the primary star is lifted and transformed from its initially spherical shape into a torus-like structure. We further find the material to be rotationally supported (i.e., circularized, and Keplerian motion) and still to be gravitationally bound. Hence, the basic requirements for the remaining envelope material to be a disk are already fulfilled \citep[e.g.,][]{pringle1981a} and in the following we want to characterize the system in more detail.
The main open questions to address in this regard are thus: is the specific angular momentum distribution such that the system can maintain a quasi-steady state, does the material develop HSE in the vertical direction, and does the system develop a (partially) evacuated cavity around the core binary.

Given the presence of the magnetized outflows in our simulations (Sect.~\ref{sec:oe}), the ejection of material perpendicular to the orbital plane (Fig.~\ref{fig:overview_disk}a) can act like a \qq{pressure valve} \citep{chamandy2018a} and prevent the virialization of the gas in the bound material around the core-binary. Therefore, the system cannot provide pressure support and, thus, disable radially inward mass migration. This already has become evident in the form of radially inward propagating material in the vicinity of the binary (Sect.~\ref{sec:morpho}) and we later follow up on this idea and measure the mass flux toward the central binary.   

Regarding the development of a quasi-steady state and vertical HSE, we face significant difficulties with the confirmation of the establishment of a CBD in our complex 3D simulations. Especially in the simulation involving the NS companion, the orbital precession further complicates matters (Sect.~\ref{sec:oe}) and we restrict our analysis to the simulation with the BH companion in the following.
There, the density as well as the specific entropy in the toroidal structure (Fig.~\ref{fig:overview_disk}b, or Movie M1 in Table~\ref{tab:movietable}) follow a clear cylindrical symmetry (except for local fluctuations). The material is perturbed in both quantities by the spiral shocks generated through gravitational torque exerted from the core-binary onto the gas in its immediate vicinity. The vertical ($z$-direction) distribution of the density and specific entropy point (Fig.~\ref{fig:overview_disk}a and c) toward a convectivly stable, Gaussian-like profile as expected for disks in vertical HSE \citep[e.g.,][]{pringle2007a}.

\begin{figure}
    \centering
    \resizebox{\linewidth}{!}{ 
    \includegraphics[width=\linewidth]{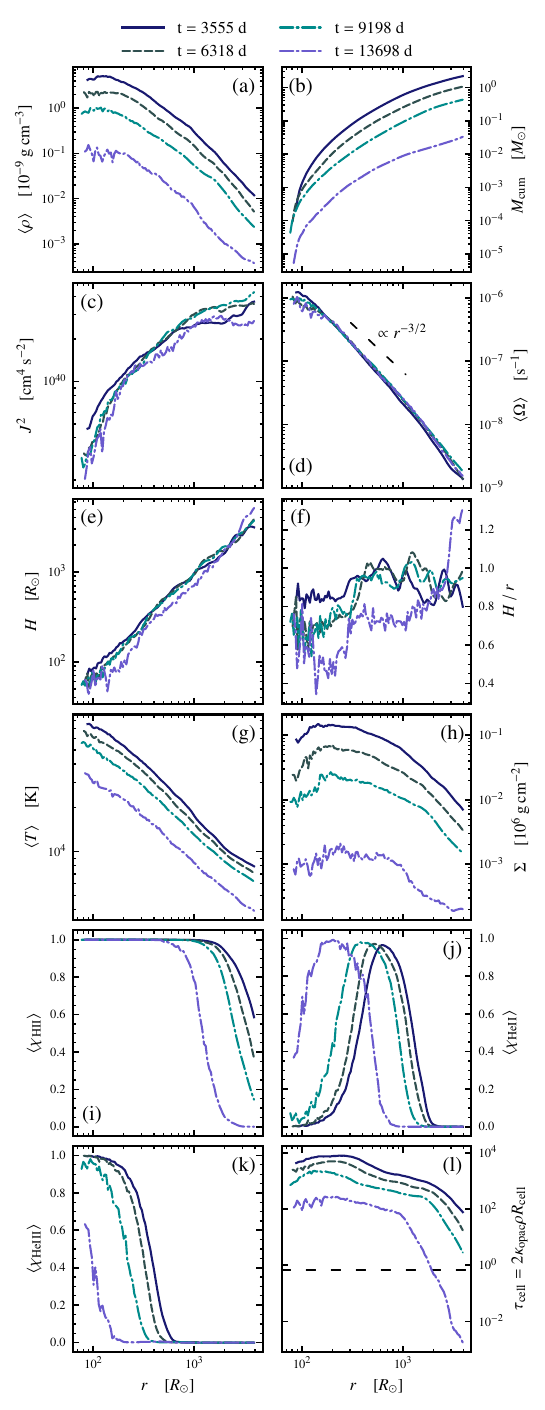}}
    \caption{Time evolution of averaged density $\langle \rho \rangle$ in (a), cumulative mass $M_\mathrm{cum.}$ in (b), specific angular momentum $J^2$ in (c), angular velocity $\langle \Omega \rangle$ in (d), isothermal pressure scale height $H$ in (e), the aspect ratio $H/r$ in (f), the temperature $T$ in (g), the column density $\Sigma$ in (h), the ionization fraction of $\mathrm{H_{\RNum{2}}}$ ($\chi_\mathrm{H_{\RNum{2}}}$), $\mathrm{He_{\RNum{2}}}$ ($\chi_\mathrm{He_{\RNum{2}}}$), and $\mathrm{H_{\RNum{3}}}$ ($\chi_\mathrm{He_{\RNum{3}}}$) in (i), (j), and (k). Last, the average optical depth of cells $ \tau_\mathrm{cell}$ is shown in (l). The averages are taken over the range of $r\,{\in}\, [2 a(t),\, 4{\times} 10^{3}\, \rsol]$ (Sect.~\ref{sec:CBD}).}
    \label{fig:disk_average_quants}
\end{figure}

To check the temporal persistence of the disk structure, we show the time evolution of the radial distribution of quantities evaluated on cylindrical\footnote{The choice for taking cylindrical averages instead of spherical averages is motivated by the weak azimuthal dependency in the bound material where the motion of the gas is circularized (e.g., Fig.~\ref{fig:disk_summary}). In fact, accounting for spherical, mass-weighted averages as, for example, in \citet{narayan1995a} leads only to small deviations, and we therefore adhere to cylindrical symmetry for simplicity.} shells in 
Fig.~\ref{fig:disk_average_quants}. For all taken averages within a cylindrical shell, we use mass-weighted averages (i.e., $\langle Q \rangle \, {=}\, \int Q \rho\, \mathrm{dV} / \int \rho \, \mathrm{dV}$ for any Q). 
Furthermore, we consider $2{\times} a(t)$ as the inner border \citep[e.g.,][]{munoz2020a,siwek2022a,siwek2023a}, while the outer boundary for our averaging is kept constant at $R_\mathrm{out} \,{=}\, 4000 \, \rsol$.
The choice for the outer radius is motivated by substantial loss of material radially outward in the orbital plane at around $R_\mathrm{out} \,{\gtrsim}\, 4000 \, \rsol$ (see streamlines in Fig.~\ref{fig:disk_summary}g, h), indicating the presence of radial winds (Fig.~\ref{fig:disk_average_quants}h) that further strip the envelope material as briefly described in Sect.~\ref{sec:morpho}.
In the first four panels of Fig.~\ref{fig:disk_average_quants}, the temporal evolution of the average density $\langle \rho \rangle$ (Fig.~\ref{fig:disk_average_quants}a), the cumulative mass $M_\mathrm{cum}$ (Fig.~\ref{fig:disk_average_quants}b), the square of the specific angular momentum $J^2$ (Fig.~\ref{fig:disk_average_quants}c) and the average angular velocity $\langle \Omega \rangle$ are shown. Unsurprisingly, the density profile in Fig.~\ref{fig:disk_average_quants}a decreases over time as the bound material is progressively ejected (Fig.~\ref{fig:disk_average_quants}b). The squared specific angular momentum (Fig.~\ref{fig:disk_average_quants}c) strictly increases along the radial direction, hence, pointing toward a Rayleigh-stable configuration \citep[$\partial J^2 {/} \partial r \,{>}\,0$, e.g.,][]{pringle2007a} and the angular velocity (Fig.~\ref{fig:disk_average_quants}d) reflects the behavior of a pressure-supported sub-Keplerian disk \citep[e.g.,][we find $\Omega \propto r^{-1.86}$]{adachi1976a, weidenschilling1977a}. Hence, the system obeys the dynamic properties of a CBD, but is experiencing mass and angular momentum loss over time.

We further try to explore characteristic properties of a CBD in panels (e) to (h) of Fig.~\ref{fig:disk_average_quants}, such as the isothermal pressure scale height $H$, the aspect ratio \inlineeq{\theta}{\equiv}{H/r}{}, the average temperature $\langle T \rangle$ and the vertically integrated column density $\Sigma \,{=}\, \int\rho\,\mathrm{d}z$. The pressure scale height in Fig.~\ref{fig:disk_average_quants}e is measured by fitting the density distribution along the vertical direction with a Gaussian profile (e.g., Fig.~\ref{fig:overview_disk}c). We find that $H$ remains approximately constant throughout the simulation. The vertical scale height is resolved with approximately 10 (3) to 60 (28) cells per $H$ at $t \,{=}\, 3555 \, \mathrm{d}$ ($t \,{=}\, 13698 \, \mathrm{d}$) for the inner and outer boundary, respectively. The aspect ratio (Fig.~\ref{fig:overview_disk}f) is fluctuating around \inlineeq{}{\approx}{0.8}{} and thus confirming the CBD to be thick. Similar to the density, the temperature (Fig.~\ref{fig:disk_average_quants}g) and column density (Fig.~\ref{fig:disk_average_quants}h) decrease with time.

\begin{figure}[h!]
    \includegraphics[width=\linewidth]{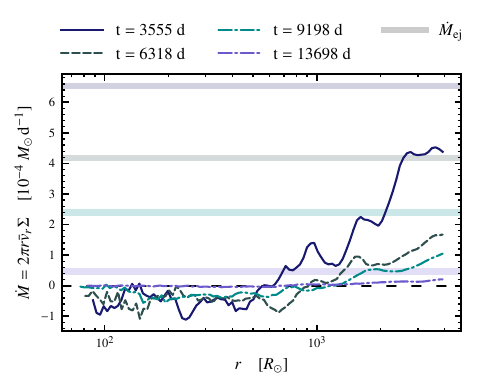}
    \caption{Radial distribution of mass transport rates $\dot{M} = 2\pi r \bar{v}_r \Sigma$ similar to Fig.~\ref{fig:disk_average_quants}. The faint bands over the entire radial domain in the background correspond to the overall envelope ejection rate $\dot{M}_\mathrm{ej}$ of the system at the corresponding time with the matching color.}
    \label{fig:M_dot}
\end{figure}

With these quantities at hand, we can now further measure the mass transport rates through the different ejection channels. Given the cylindrical symmetry of the system, the mass flux can be approximated by the average transport of cylindrical shells $\dot{M} \, {=}\, 2\pi r \bar{v}_r \Sigma$ with $\bar{v}_r$ being the average cylindrical radial velocity. With this definition, negative (positive) values correspond to radial inward (outward) propagation of the corresponding shell. We note that with this definition, the outward mass transport rates may be underestimated because the recombination energy is locally thermalized, and the resulting velocity field tends toward spherical symmetry, depending on the local pressure gradient (see Fig.~\ref{fig:disk_summary}c and d).
The resulting radial profile of the transport rates is shown in Fig.~\ref{fig:M_dot}. The mass flux is dominantly inward for radii \inlineeq{r}{\lesssim}{10^3}{\rsol} and points outward for lager radii, reflecting our findings from Sect.~\ref{sec:morpho}, where we concluded that the bound torus is eroded by radial winds and accretion toward the core binary, which, before reaching it, is channeled into a jet-like outflow. 

The negative mass transport rates in the inner part of the disk (out to about $600 \, \rsol$ in Fig.~\ref{fig:M_dot}, which we will in the following refer to as inflow rates) appear to be relatively constant throughout the simulation. They are on the order of $\dot{M} \,{\sim}\, - 10^{-5}\, \msol\, \mathrm{d^{-1}}$ and decrease significantly only for very late times (for \inlineeq{t}{=}{13,698}{\mathrm{d}}, where we measure $\dot{M} \,{\sim}\, - 10^{-6}\, \msol\, \mathrm{d^{-1}}$, see Fig.~\ref{fig:M_dot}). The notion of inflow rates over the disk is supported by computing the $\alpha$-viscosity given by the mass weighted average \citep{balbus1998a}, 
\begin{align}
    \alpha = \left \langle \frac{\delta v_r \delta v_\phi}{c_s^2} - \frac{B_rB_\phi}{4\pi \rho c_s^2} \right \rangle, \label{eq:alpha_visc}
\end{align}
where $\delta v_r \,{=}\, v_r$ and $\delta v_\phi \,{=}\, v_\phi - \sqrt{G(M_\mathrm{core}+M_2)/r}$ are the radial and azimuthal velocity perturbations, $c_s$ is the sound speed, $B_r$ and $B_\phi$ are the radial and azimuthal magnetic field components. This yields average $\alpha$-viscosity values in the inner (i.e., well within the hydrogen recombination front; $R\, {\lesssim} \, 600 \, R_\odot$) parts of the bound material of ${\approx}\, 0.06$ to $0.13$. This is in good agreement with the expectations from local magneto-rotational instability (MRI) simulations with $\alpha$-values on the order of $10^{-2}\,\textit{--}\, 10^{-1}$ \citep[e.g.,][]{davis2010a}. Furthermore, we can reproduce the inflow rates measured in our simulation by an $\alpha$-disc model \citep[][]{shakura1973a},
\begin{align}
    \dot{M}_\mathrm{visc} &= -3 \pi \alpha r^2 \Omega \theta^2 \Sigma \approx -(10^{-4}\,\textit{--}\, 10^{-6}) \, \msol \, \mathrm{d^{-1}},
\end{align}
for $\alpha \,{\approx}\, 10^{-1}$ at the inner disk edge (i.e., $r\,{\approx}\, 80 \, \rsol$) with the local angular velocity $\Omega \,{\approx}\, 10^{-6}\, \mathrm{s^{-1}}$, the aspect ratio $\theta\,{\approx}\, 0.8$ (Fig.~\ref{fig:disk_average_quants}f) and the integrated column density $\Sigma \,{\approx}\, 8{\times}10^{4} \, \mathrm{g\, cm^{-2}}$ ($\Sigma \,{\approx}\, 8{\times}10^{2} \, \mathrm{g\, cm^{-2}}$) for $t\,{=}\,3555 \, \mathrm{d}$ ($t\,{=}\,13,698 \, \mathrm{d}$). Most (if not all) of this inward-transported material, however, is found to be launched into the magnetically driven outflow mentioned in Sect.~\ref{sec:oe} (Vetter et al.~in prep.). It will not participate in altering the orbital separation via angular momentum accretion onto the engulfed core binary system \citep[e.g.,][]{gagnier2023a, wei2023a, siwek2022a}.

The mass transport in the outer regions (i.e., beyond the hydrogen recombination front at maximal radii in Fig.~\ref{fig:M_dot}) exceeds the inflow rates in the inner parts by approximately one order of magnitude (\inlineeq{\dot{M}}{\sim}{10^{-4}\,\textit{--}\,  10^{-5}}{\msol\,d^{-1}}) and thus is expected to dominate the ejection rates in the entire system. In fact, we find good agreement in the overall mass ejection rates in the system $\dot{M}_\mathrm{ej}$ (Fig.~\ref{fig:M_dot}). 

The outer transport region coincides with the recombination front of hydrogen and helium (see Fig.~\ref{fig:disk_average_quants}i--k, where we show the ionization fraction of $\mathrm{H_{\RNum{2}}}$, $\mathrm{He_{\RNum{2}}}$ and $\mathrm{He_{\RNum{3}}}$). Assuming full conversion of released recombination energy to kinetic energy, the expected velocities of the ejecta escaping the gravitational pull of the core binary (${\sim}\, 47 \, \mathrm{km \, s^{-1}}$ for helium and hydrogen and ${\sim}\, 38 \, \mathrm{km \, s^{-1}}$ for only hydrogen recombination)\footnote{In this approximation, we use $v\,{=}\,[(Y/2\Delta E_\mathrm{He_{III} \rightarrow He_{I}} + 2X\Delta E_\mathrm{H_{II} \rightarrow H_{I}})m_p^{-1} + \Delta e_\mathrm{grav}]^{1/2}$, with $Y=0.3$ and $X=0.7$ the helium and hydrogen mass fractions, $m_p$ the proton mass and $\Delta E_\mathrm{He_{III} \rightarrow He_{I}}$ as well as $\Delta E_\mathrm{H_{II} \rightarrow H_{I}}$ the recombination energy of helium and hydrogen, respectively. The specific relative gravitational energy $\Delta e_\mathrm{grav} \,{=}\, - 0.5G(M_2 + M_\mathrm{core})R^{-1}$ is computed with the primary-star-core mass $M_\mathrm{core} \,{\approx}\, 2.97 \, \msol$, the BH companion mass $M_2 \,{\approx}\, 5 \, \msol$ and the position of hydrogen recombination ${R \,{\approx}\, 10^{3} \rsol}$.}\label{footnote:energy_argument} approximately match the characteristic outflow velocities observed in our simulation of ${\sim}\, 42\, \mathrm{km\, s^{-1}}$ (Fig.~\ref{fig:disk_summary}g). This supports the argument for a recombination-driven radial wind.
As the material is progressively ejected, the remaining bound material cools adiabatically (Fig.~\ref{fig:disk_average_quants}g), leading to inwardly propagating recombination fronts (Fig.~\ref{fig:disk_average_quants}i, j, k) and finally to the ejection of the material according to Eq.~(\ref{eq:e_kin_crit}). 

Given the sharp decline in optical depth of the cells $\tau_\mathrm{cell} \,{=}\, 2\kappa_\mathrm{opac} \rho R_\mathrm{cell}$ (with $\kappa_\mathrm{opac}$ being the opacity) closely behind the hydrogen recombination front (see Fig.~\ref{fig:disk_average_quants}l), the ejection of the bound material in the radial winds could be influenced by cooling. The adiabatic assumption in our simulations (i.e., local thermalization of the released recombination energy) could thus overestimate the amount of ejected material in the radial winds through hydrogen recombination. In this case, more material would be expected to remain bound in the CBD and further evolve toward a thin disk structure.

%--------------------------------------------------------------------
\subsection{Common-envelope ejection efficiency}\label{sec:results_alpha_CE}
In the following, we are trying to compare the expectations from the $\alpha$-formalism to the final orbital separations found in our simulations. Applying the alpha-energy formalism \citep{webbink1984a},
\begin{align}
    \alpha_\mathrm{CE} = \frac{E_\mathrm{bind}}{\Delta E_\mathrm{orb}}, \label{eq:alpha_ce}
\end{align}
where $\alpha_\mathrm{CE}$ is the envelope ejection efficiency and $E_\mathrm{bind}$ the binding energy of the envelope \citep[][]{webbink1984a,moreno2022a}, 
\begin{align}
    E_\mathrm{bind} &= - \int_{M_\mathrm{core}}^{M_\mathrm{surf}} \frac{Gm}{r} \, \mathrm{d} m + \alpha_\mathrm{th} \int_{M_\mathrm{core}}^{M_\mathrm{surf}} u \, \mathrm{d} m \label{eq:binding_energy}\\
    &\equiv -\frac{GM_1M_\mathrm{env}}{\lambda R_1}, \label{eq:binding_energy_lambda}
\end{align}
and $\Delta E_\mathrm{orb}$ the released orbital energy,
\begin{align}
    \Delta E_\mathrm{orb} = - \frac{G M_\mathrm{core} M_\mathrm{2}}{2a_\mathrm{f}} + \frac{G M_\mathrm{1} M_\mathrm{2}}{2a_\mathrm{i}},\label{eq:orbital_energy}
\end{align}
we can calculate the final post-CE orbital separation for $\alpha_\mathrm{CE}$ and $\alpha_\mathrm{th}$ chosen to be unity. 
Here, $M_\mathrm{core}$ is the primary-core mass, $u$ the internal energy (containing recombination energy), $m$ the mass coordinate with corresponding radius $r$, and $\lambda$ as well as $\alpha_\mathrm{th}$ parameters accounting for the available binding energy of the envelope \citep{dekool1990a} and the fraction of total internal energy aiding to unbind the envelope. 
We note that the integral in Eq.~(\ref{eq:binding_energy}) is performed from an a priori unknown mass coordinate of the stellar core to the surface of the star. This intrinsically raises the problem of defining the core and, hence, the envelope of the star and is often approximated in binary evolution calculations by, for instance, the maximum compression point or the point, where the hydrogen abundance drops below \inlineeq{X}{=}{0.1}{} (see Sect.~\ref{sec:initial_setup}, \citealp{moreno2022a}).

\begin{table}[h]
    \centering
    \setlength{\tabcolsep}{6.3pt}
    \caption{Determined CE ejection efficiencies $\alpha_\mathrm{CE}$ based on the results of the 3D CE simulations.}
    \begin{tabular}{ccccccccc}
	\toprule
	& & \multicolumn{3}{c}{BH} & \multicolumn{3}{c}{NS} \\
	$\alpha_\mathrm{th}$ & $\lambda$ & \multicolumn{3}{c}{$\alpha_\mathrm{CE}$} & \multicolumn{3}{c}{$\alpha_\mathrm{CE}$} \\
    & & $\kappa=0$ & & $\kappa=1$ & $\kappa=0$ & & $\kappa=1$ \\
	\midrule
	0.0 & $0.483$ & $2.57$ & $\rightarrow$ & $1.86$ & $2.29$ & $\rightarrow$ & $2.12$\\
	0.5 & $0.763$ & $1.63$ & $ \rightarrow $ & $1.18$ & $1.45$ & $\rightarrow$ & $1.34$\\
	1.0 & $1.81$ & $0.69$ & $ \rightarrow $ & $ 0.50$ & $0.61$ & $ \rightarrow $ & $ 0.57$\\
	\bottomrule
	\end{tabular}
    \tablefoot{$\alpha_\mathrm{th}$ and $\lambda $ are defined in Eq.~(\ref{eq:binding_energy}) and (\ref{eq:binding_energy_lambda}), where the $\lambda$ values are extracted from \citet{moreno2022a} (see Sect.~\ref{sec:results_alpha_CE} for the details of the calculation).}
    \label{tab:alpha_ce}
\end{table}

The binding energy of the envelope $E_\mathrm{bind}$ is obtained via integrating Eq.~(\ref{eq:binding_energy}) over the one-dimensional \textsc{mesa} stellar profile \citep[][Fig.~1b]{moreno2022a} and assuming full accessibility of the internal energy reservoir (i.e., $\alpha_\mathrm{th}\,{=}\,1$). Here, the stellar core mass in Eq.~(\ref{eq:binding_energy}) (and consequently the envelope mass in Eq.~\ref{eq:binding_energy_lambda}) is chosen to be defined by the compression point of the star (Sect.~\ref{sec:initial_setup}). 
By rearranging Eq.~(\ref{eq:alpha_ce}) and Eq.~(\ref{eq:orbital_energy}) and further assuming $\alpha_\mathrm{CE} \,{=}\, 1$, the expected final separations are $a_\mathrm{f, \, BH} \,{\approx}\, 19.2\, \rsol$ and $a_\mathrm{f,\, NS} \,{\approx}\, 5.9\, \rsol$ for the BH and NS companion, respectively \citep[][]{moreno2022a}. These values are in contrast to the larger final separations of our CE simulations of $a_\mathrm{f, \, BH} \,{\approx}\, 37\, \rsol$ and $a_\mathrm{f, \, NS} \,{\approx}\, 14\, \rsol$. 
This discrepancy was also found in lower mass systems \citep[e.g.,][]{ohlmann2016a,iaconi2017a, sand2020a, kramer2020a}. 

This difference may be because the maximum compression point is not contained on our simulation domain, leading to a lower envelope binding energy. In the following, we use the binding energy of the envelope from the relaxed \textsc{arepo} RSG progenitor model, which is up to a factor of 6 smaller \citep[see][Fig.~1b]{moreno2022a}. It is therefore plausible for the above estimate to yield final orbital separations reduced by the same factor.
The integral is performed from the mass coordinate defined by the initial cut radius ($M_\mathrm{core} \,{=}\, 2.97 \, \msol$) to the surface of the star ($M_1 \,{=}\, 9.4 \, \msol$). 
For $\alpha_\mathrm{th} \,{=}\, 0.0, \, 0.5$ and $1.0$ we find through Eq.~(\ref{eq:binding_energy_lambda}) $\lambda \,{=}\, 0.483, \, 0.763$ and $1.810$, respectively \citep[Table~\ref{tab:alpha_ce},][]{moreno2022a}.\footnote{We account for $\alpha_\mathrm{th} \, {<}\, 1$ to judge what would happen if the recombination energy cannot be fully used due to radiation cooling. In our model we always have the full internal energy available, hence, $\alpha_\mathrm{th} \, {=} \, 1$.}

Combining Eq.~(\ref{eq:binding_energy_lambda}) and Eq.~(\ref{eq:orbital_energy}) yields
\begin{align}
    \alpha_\mathrm{CE} = \frac{GM_\mathrm{1}M_\mathrm{env}}{\lambda R_\mathrm{1}} \left [ \frac{GM_\mathrm{core}M_\mathrm{2}}{2a_\mathrm{f}} - \frac{GM_\mathrm{1}M_\mathrm{2}}{2a_\mathrm{i}}\right ]^{-1}, \label{eq:alpha_ce_2}
\end{align}
which results for our measured final orbital separations (Sect.~\ref{sec:oe}) in $\alpha_\mathrm{CE, \, BH} \,{=}\, 0.5\, \textrm{--} \, 1.86$ and $\alpha_\mathrm{CE, \, NS} \,{=}\, 0.57\, \textrm{--} \, 2.12$, depending on the choice of $\alpha_\mathrm{th}$ (see Table~\ref{tab:alpha_ce}). 

\begin{figure}
    \centering
    \includegraphics[width=\linewidth]{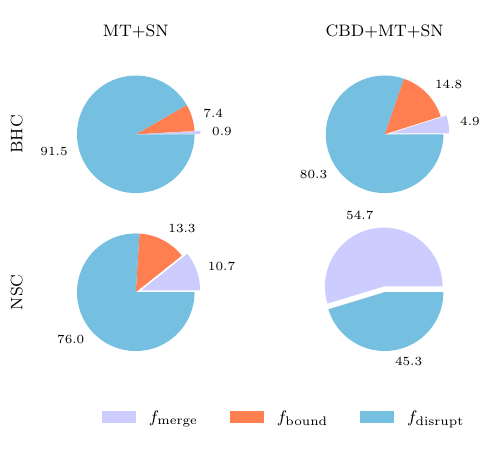}
    \caption{Fractions in percent of systems that are considered to merge within the Hubble time $f_\mathrm{merge}$ (light purple), expected to stay in a bound binary configuration and are still too wide to merge within a Hubble time $f_\mathrm{bound}$ (light blue) and to be disrupted $f_\mathrm{disrupt}$ (orange) by the natal NS kick. Considered are the following evolutionary pathways after the CE event: Either the post-CE binary first experiences an MT episode followed by the SN (MT+SN) or the CBD interaction and an MT episode transitions into the SN (CBD+MT+SN). All events are modeled separately, and the assumptions for the calculations are provided in Sect.~\ref{sec:final_fate}.}
    \label{fig:final_fate}
\end{figure}

%--------------------------------------------------------------------
\subsection{Final fate of the systems} \label{sec:final_fate}
Given the final orbital separations of $a_\mathrm{f, \, BH} \,{\approx}\, 37 \, \rsol$ and $a_\mathrm{f,\, NS} \,{\approx}\, 14 \, \rsol$, and eccentricities of $e_\mathrm{BH} \,{=}\, 0.12$ and $e_\mathrm{NS} \,{=}\, 0.055$ for the BH and NS companion, respectively, we find merging times of $T_\mathrm{merge,\, BH} \,{=}\, 2320 \, \mathrm{Gyr}$ and $T_\mathrm{merge,\, NS} \,{=}\, 325 \, \mathrm{Gyr}$ by GW emission \citep{peters1964a}. Hence, the systems are not expected to merge within a Hubble time of $13.787 \, \mathrm{Gyr}$ \citep{planckcollab2020a}.

The extended lifetime of the post-CE binaries set by the -- compared with the CE phase -- inefficient inspiralling process due to GW emission and the remaining lifetime of the RSG remnant of $6{\times}10^{4} \, \mathrm{yr}$ \citep{moreno2022a} opens up several evolutionary pathways, that could potentially alter their fates tremendously. These include a second stable MT phase initiated by the re-expansion of the primary core \citep[e.g.,][]{tauris2017a, woosley2019a, laplace2020b, jiang2021a, vigna2022a,wei2023a}, the interaction of the engulfed post-CE binary with the left-over bound material in the orbital plane \citep[e.g.,][]{gagnier2023a, tuna2023a,wei2023a} and the SN event of the primary core forming a NS (Sect.~\ref{sec:initial_setup}) and a connected NS kick \citep[e.g.,][]{brandt1995a,wei2023a}. The results of \citet{wei2023a} underpin the inevitability of the MT as well as the SN event, narrowing down the possible pathways.

To account for the MT, we closely follow the analysis of \citet{moreno2022a} and consider the companions to be genuine compact objects with Eddington-limited accretion.\footnote{For the system involving the NS companion, we consider $\dot{M}_\mathrm{edd}^\mathrm{NS}\,{=}\, 4\pi c R\, /\,\kappa$, while for the BH companion scenario, we adhere to $\dot{M}_\mathrm{edd}^\mathrm{BH}\,{=}\, 4\pi G M_\mathrm{BH}\, /\,\kappa c \eta$. Here, $c$ is the speed of light and $\kappa \,{=}\, 0.2(1+X)\,  \mathrm{cm^{2}g^{-1}}$ the electron scattering opacity with hydrogen. In our model, $X\,{=}\,0.7$ and we assume $\eta \,{=}\, 0.07$ \citep{podsiadlowski2003a}.} This effectively leads to almost completely nonconservative MT.
For quantifying the evolution of the orbital separations, we utilize Eq.~(3) of \citet{tauris1996a} and, for example, Eq.~(8) of \citet{podsiadlowski2002a}. The impact of an MT event on the final orbital separations and the merging time, given our assumptions, is listed in the second row of Table~\ref{tab:final_fate}. Notably, the systems involving a NS show a decrease in orbital separations, whereas the systems with the BH companions expand.

\begin{table}
    \setlength{\tabcolsep}{1.pt}
	\caption{Change of the final orbital separation $a_\mathrm{f}$ relative to the post-CE distance of the binary $a_\mathrm{i}$ and ratio of merging time by GW emission $T_\mathrm{merge}$ and the Hubble time ($T_\mathrm{Hubble}\,{=}\,13.787 \, \mathrm{Gyr}$, \citealp{planckcollab2020a}) for possible post-CE evolutionary pathways.}
	\label{tab:final_fate}
	\begin{tabular*}{\linewidth}{l|cc|cc}
	\toprule
     &\multicolumn{2}{c}{BH} & \multicolumn{2}{c}{NS}\\
     & $a_\mathrm{f}\, / \, a_\mathrm{i}$ & $T_\mathrm{merge}\, / \, T_\mathrm{Hubble}$ & $a_\mathrm{f}\, / \, a_\mathrm{i}$ & $T_\mathrm{merge}\, / \, T_\mathrm{Hubble}$ \\
	\midrule
    post-CE       & $1.0$  & $1.7{\times}10^{2}$ & $1.0$ & $2.4{\times}10^{1}$ \\
    MT            & $1.77$ & $3.1{\times}10^{3}$ & $0.72$ & $1.3{\times}10^{1}$ \\
    CBD           & $0.45$ & $2.7{\times}10^{0}$ & $0.25$ & $3.1{\times}10^{-2}$ \\
    SN (no kicks) & $1.31$ & $1.1{\times}10^{3}$ & $2.19$ & $5.1{\times}10^{2}$  \\
    CBD+MT        & $0.72$ & $3.0{\times}10^{1}$ & $0.12$ & $3.6{\times}10^{-3}$ \\
	\bottomrule
	\end{tabular*}
    \tablefoot{The presented scenarios are: MT episode, CBD interaction and SN event without kick (SN) as in \citet{moreno2022a} (see Sect.~\ref{sec:final_fate}) for example. We also show various concatenations of evolutionary events and chain their labels together with a \qq{+.} We note that for the chained events, the output of the previous event is taken into account. We refer to the text in Sect.~\ref{sec:final_fate} for the detailed parameter considerations for the individual calculations.}
\end{table}

The interaction of the binary with the left-over bound material -- unaccounted for in the preceding work of \citet[Sect.~4]{moreno2022a} -- can drastically alter the orbital separation \citep[e.g.,][]{shi2012a, tiede2020a, siwek2022a,gagnier2023a, tuna2023a,wei2023a} and therefore the final fate of the binary on timescales of $10^3 \, \textrm{--} \, 10^5\, \mathrm{yr}$ (when photo-evaporation is expected to become important) even before the MT event takes place. 
In fact, the presence of a thick CBD may even harden massive binaries down to distances, where the emission of GWs becomes important or lead to a merger before the SN of the remnant He core \citep{dittmann2022a,wei2023a}, depending on the mass and the lifetime of the disk. 
In the following, we suppose that recombination-driven ejection is not perfectly efficient due to nonadiabatic effects. In this scenario, the remaining material can form a long-lived CBD after the dynamic CE phase that interacts with the engulfed core binary on timescales much longer than the simulated duration (Sect.~\ref{sec:oe}).
The orbital change due to CBD interaction in our systems can then be described by parameterized torque prescription provided in \citet{tuna2023a}, 
\begin{align}
    \frac{a(t)}{a_0} = \left (1 - \frac{M_\mathrm{acc}}{M_\mathrm{binary,0}} \right)^{2\chi (1+q)^2/ q -3}, \label{eq:a_accretion} 
\end{align}
where $a_0$ is the initial orbital separation, $M_\mathrm{acc}(t)$ the accreted mass, $M_\mathrm{binary,0} \,{=}\, M_\mathrm{core} + M_\mathrm{2}$ the initial binary mass, $q \,{=}\, M_\mathrm{2}/M_\mathrm{core}$ the binary mass ratio, and $\chi$ a dimensionless parameter quantifying the torque on the binary per unit accreted mass. We further express the accreted mass by the inflow rate from Sect.~\ref{sec:CBD} (Fig.~\ref{fig:M_dot}) such that $M_\mathrm{acc}\, {\approx}\, \dot{M}_\mathrm{acc} t_\mathrm{disk}$. Here, we assume that the accretion rates onto the core binary are $1 \,\mathrm{\%}$ of the inflow rates (Sect.~\ref{sec:CBD}, given that most of the inflowing material is ejected through the magnetically driven outflow) and that the expected lifetime of the disk $t_\mathrm{disk} \,{=}\, 10^{4}\, \mathrm{yr}$ is set by photo-evaporation \citep[e.g.,][]{tuna2023a}. In this way, the total accreted mass amounts to $0.73 \, \msol$ and the only unknown variable left is $\chi$. 
Since $\chi$ depends on the mass accretion onto the central binary (preventing us from computing it directly) and the literature is currently lacking any precise values, we follow the results of \citet{siwek2023a,siwek2022a}, who found that systems with $q \,{\gtrsim}\, 0.3$ tend to shrinkage, and assume $\chi \,{=}\, - 2\chi_\mathrm{crit} \,{=}\, -3 q/(1+q)^2$ as an approximation for all our simulations ($\chi_\mathrm{crit}$ is the critical value dividing widening for $\chi > \chi_\mathrm{crit}$ and broadening for $\chi < \chi_\mathrm{crit}$ of the binary orbit, \citealp{tuna2023a}). It must be noted that these estimates of the further evolution are rather a qualitative approximation, given the large uncertainties. The impact on the final distance and, thus, the merging times of the systems are summarized in the third row in Table~\ref{tab:final_fate}, where we assumed an equilibrium eccentricity of $e\,{=}\,0.5$ \citep{siwek2023a}.

For the SN kick, we adhere to the analysis in \citet[Sect.~2.1]{brandt1995a} assuming a Maxwellian distribution of kick velocities with $\sigma \,{=}\, 265 \, \mathrm{km \, s^{-1}}$ \citep{hobbs2005a} and the formation of a $1.4\, \msol$ NS.\footnote{It should be pointed out that the progenitor of the second NS is very likely to be ultra-stripped after experiencing case BB MT. For simplicity, we do not reduce the natal kicks of ultra-stripped SNe, which would greatly enhance the chances for the orbit to remain intact \citep[e.g.,][]{vigna2018a}.} Under these assumptions, however, the natal NS kick with the kick parameters as described above alters the final fate of the core binary system, as illustrated in Fig.~\ref{fig:final_fate}. In this figure, the probability of the system to merge via GW emission (fraction $f_\mathrm{merge}$) is shown in purple, the fraction of bound (but not merging) systems $f_\mathrm{bound}$ is shown in orange, and the fraction of systems disrupted by the SN kick $f_\mathrm{disrupt}$ is indicated in blue. As possible pathways for the systems, we consider the following two scenarios:
\begin{itemize}
    \item[(i)]After the CE evolution, an MT phase occurs, followed by the SN explosion of the primary core, excluding interaction with the CBD. The MT induces orbital widening in binaries with a BH companion, while for systems involving a NS companion, the separation decreases (Table~\ref{tab:final_fate}). The merger rates for scenarios involving a BH companion amount to $0.9\, \mathrm{\%}$ and $10.7\,\mathrm{\%}$ for those with a NS companion. Compared to \citet{moreno2022a}, the merger rates increase by $0.3\,\mathrm{\% pt}$ and $3.0\,\mathrm{\% pt}$ for the BH and NS scenarios, respectively.   
    \item[(ii)]We consider the CBD interaction and an MT episode ending with the SN event. Physically, the CBD and MT event can take place simultaneously \citep{wei2023a, tuna2023a}, but we choose to model them as subsequent phases. The CBD interaction then first hardens all separations and the MT further tightens the systems involving NS companions while widening those with BH companions (Table~\ref{tab:final_fate}). As a result, the systems with NS companions increase their merger rates once more to $54.7\, \mathrm{\%}$, while the merger fraction in the systems containing a BH enhances the merger fraction to $4.9\, \mathrm{\%}$.
    \end{itemize}

In summary, the interaction between the core binary and a CBD can significantly tighten the binaries and increase the likelihood of a merger event, but the individual results should be interpreted cautiously due to the considerable uncertainties arising from our simplified assumptions. A comprehensive exploration of the parameter space and the development of more sophisticated models, such as those proposed by, for example, ~\citet{wei2023a}, are essential for obtaining more detailed insights.

%%%%%%%%%%%%%%%%%%%%%%%%%%%%%%%%%%%%%%%%%%%%%%%%%%%%%%%%%%%%%%%%%%%%%
\section{Discussion}\label{sec:discussion}
%--------------------------------------------------------------------
\subsection{Evolution of the orbital separations}\label{sec:discussion_oe}
The tension between the orbital separations resulting from CE interactions according to expectations of the $\alpha$ formalism and the outcomes of our simulations (Sect.~\ref{sec:results_alpha_CE}) is not unique to our modeled scenarios. In fact, the findings in lower mass systems \citep[e.g.,][]{passy2012a, ricker2012a, ohlmann2016a, iaconi2017a, prust2019a, sand2020a, reichardt2020a, ondratschek2022a} or more massive systems \citep[e.g.,][]{moreno2022a,lau2022a} reach similar conclusions. 
In contrast to the energy prescription of the CE evolution and in agreement with, for instance, \citet{sand2020a}, after the initial perturbation of the envelope by the rapid spiral-in of the companion, the envelope ejection induced by the release of recombination energy barely affects the orbital separations (Sect.~\ref{sec:oe}). 
Consequently, our final orbital distances significantly exceed the separations expected from the energy formalism (Sect.~\ref{sec:results_alpha_CE}). Even under full envelope ejection, the distances only marginally decrease compared to \citet{moreno2022a}.

Physically, the strongly decreased orbital contraction rates after the spiral-in of the core binary are caused by the ceasing of the dynamical drag force \citep[e.g.,][]{ostriker1999a}, which is induced by the rapidly decreasing density of the gas in the vicinity of the core binary system, together with the subsonic and partially corotating motion of this gas \citep[e.g.,][]{moreno2022a,roepke2022a}. The reduced drag is then operating on the binary until the system dynamically equilibrates, and the further evolution is set by processes on longer timescales \citep[e.g.,][]{ivanova2013a,hirai2022a}. Given the occurrence of a CBD directly after the rapid in-spiral, this picture must be complemented by gravitational torques (or other stress-related angular momentum extraction processes) and the accretion of mass and angular momentum onto the primary core and companion, which could affect the post-spiral-in contraction. These processes could even proceed on longer timescales and are shown to impact the further evolution of the core binary significantly \citep[][]{wei2023a}.

Our simulations with $\kappa\, {=}\, 1$ show good agreement with the simulations involving $\kappa \, {=} \, 0$ (Sect.~\ref{sec:oe} and Appendix~\ref{sec:appendix_refinement}). These, in turn, show convergence in the (averaged) orbital evolution and envelope ejection with regard to the initial cut radius according to the resolution study presented in Fig.~9 of \citet{moreno2022a}.\footnote{Convergence in final orbital separations and envelope ejection for CE simulations conducted with \textsc{arepo} is also shown for different CE systems for a sufficiently low cut radius and high enough resolution per softening length \citep[e.g.,][]{ohlmann2016a, kramer2020a}} It therefore appears plausible to conclude that the found (averaged) final orbital separations are indeed not only an upper limit but reflect the expected outcome given the considered timescales and physical model. The discrepancy to the $\alpha$-CE formalism can thus only be mitigated by reporting updated envelope ejection efficiencies (e.g., Sect.~\ref{sec:results_alpha_CE}).

Most numerical and physical uncertainties concerning the orbital separations are related to approximating the primary core and companion as gravity-only point particles (Sect.~\ref{sec:initial_setup}). We focus here on arguments that were not accounted for in previous publications, such as the adaptive reduction of the softening length \citep{moreno2022a} and the reaction of the primary core to envelope loss (e.g., \citealp{wei2023a}, see also \citealp{bronner2023a}), and refer to these publications for a more detailed discussion.

As stated in Sect.~\ref{sec:initial_setup}, with the gravity-only point particle approximation, we neglect any accretion of angular momentum and mass onto the primary core and the companion during our simulation. 
However, this is expected to leave only a weak impact on the orbits over the dynamical CE interaction.  
Quantitatively, this can be shown by applying Eq.~(\ref{eq:a_accretion}) and assuming the extreme case that the inner binary can accrete mass an angular momentum on a similar rate than the inflow rates. Assuming further, that this process proceeded from the onset of the magnetized outflows to the end of our simulations (i.e., \ $M_\mathrm{acc}\,{\approx}\, 2{\times} 10^{-5}\, \msol \mathrm{d^{-1}} [t_\mathrm{end} - t_\mathrm{0}]$, where $t_\mathrm{end}$ is the total simulated time and $t_0$ is set by the onset of the magnetized outflows) and $\chi\,{=}\,-3 q/(1+q)^2$ as in Sect.~\ref{sec:final_fate}, this results in deviations from the final orbital separation on the order of $0.1\, \mathrm{\%}$. Hence, the accretion onto the central binary seems to be unimportant for the overall evolution during the dynamical CE interaction.

We find changes in the eccentricity during the late stages of our simulations (as can be qualitatively inferred from the fluctuations in the orbital separations in Fig.~\ref{fig:dyn_evolution} and seen in Fig.~\ref{fig:appendixB_eccentricity}).
As already discussed in \citet{moreno2022a}, our simulations do not converge in eccentricity, and the reported values in Sec.~\ref{sec:final_fate} must be taken with caution. 
The arising changes seem to be inherently connected to the point-particle approximation of the companion and primary star core, but there is currently no coherent explanation for this behavior. One potential cause of these changes might be the adaptive reduction of the softening length, which may introduce perturbations to the HSE (if present). 
However, the abrupt changes in the eccentricities do not correlate with the reduction steps in the softening length (compare Fig.~\ref{fig:dyn_evolution} and Fig.~\ref{fig:appendixB_comp_costs}a and \ref{fig:appendixB_comp_costs}b) and the total energy is conserved down to a few per mil throughout our simulations (relative to the initial total energy in the system the deviations amount to $0.11\, \%$ for the BH and $ 0.16\, \%$ for the NS companion case, respectively).
Moreover, the eccentricity changes are mostly observed in the later stages of the simulations, where the softening length either changes only smoothly or is constant. Therefore, a direct connection between the changes in eccentricity and the changes in the gravitational softening length seems unlikely. While some other numerical artifacts cannot be excluded as a reason, the changes in eccentricity might have a physical origin, for example, from the complex interaction of the bound material and the central binary, similar to the effects associated with warped disks \citep[e.g.,][]{nixon2016}.

%--------------------------------------------------------------------
\subsection{The common-envelope ejection and the disk-like bound material}\label{sec:disk_discussion}
In Sect.~\ref{sec:morpho} and Sect.~\ref{sec:CBD}, we found evidence for the presence of a thick CBD surrounding the core binary after the rapid spiral-in. The material showed characteristics of a thick, pressure-supported sub-Keplerian disk.
We further identified two envelope ejection channels in our simulations, and in contrast to the spherical expansion, the channels are tightly coupled to the CBD. The first channel is a disk wind driven by recombination and stripping the outer layers of the torus-like structure. The second channel is the ejection of material through a magnetically driven jet-like outflow in the polar directions, which is refueled by the CBD material. 

Since the first channel is driven by recombination, it depends sensitively on the coupling of the radiation field to the hydrodynamic evolution. Our simulations are fully adiabatic and assume local thermalization of recombination energy. Ultimately, this leads to the envelope ejection, as shown in Fig.~\ref{fig:dyn_evolution}, predominantly via this channel. 
Physically, however, the photosphere is expected to be near the hydrogen recombination front (Fig.~\ref{fig:disk_average_quants}) and some of the energy may be lost by radiation without performing mechanical work. Hence, the ejection efficiency by the radial winds could be overestimated. Therefore, a more advanced model incorporating radiative cooling processes is necessary to address this issue comprehensively. 

While the question of how much hydrogen recombination energy is accessible to the system is highly speculative, one could also turn the question around and ask how much of the hydrogen recombination energy is required to accelerate material at the hydrogen recombination front beyond the local escape velocity. For the BH companion, this question can be approximated by equating the specific energies $G(M_\mathrm{c}+M_2)/ 2R$ and $\Delta e_\mathrm{H_{II} \rightarrow H_{I}}$ at the hydrogen recombination front (here denoted by $R$; c.f.\ argument in footnote~3). For \inlineeq{R}{\approx}{4\,{\times}10^{3}}{\rsol} this implies that almost $21\, \mathrm{\%}$ of the recombination energy is required to eject the material according to the kinetic-energy criterion and as the recombination front recedes in the evolution, this number becomes progressively higher (e.g., at around \inlineeq{t}{\approx}{13,698}{d}, \inlineeq{R}{\approx}{10^{3}}{\rsol} and thus $83\, \mathrm{\%}$ are needed). This leads us to the conclusion that, under the influence of cooling, some of the disk material can remain bound and form a long-lived CBD.  
Assuming further that the disk winds can indeed be lowered or even ceased by cooling (depending on the cooling timescale compared to the ejection timescale), the mass transport inward can still be maintained by the persistence of turbulent transport mechanisms (like the saturated MRI phase) or angular momentum extraction by gravitational torques inflicted by the engulfed binaries. In this scenario, the second mass-loss channel through the jet-like polar outflow could become the only way for the system to eject material.

While we analyze the properties of the magnetically driven outflow in detail in Vetter el al.~(in prep.), we can already conclude from our analysis of the inflow rates and the disk wind in Sect.~\ref{sec:CBD} that the envelope mass lost through this channel has to be subdominant during our simulations. 
So far, it is not resolved by which mechanisms angular momentum is transported, and a more elaborate analysis \citep[as attempted in e.g.,][]{gagnier2023a,gagnier2024a} is required to address this question. One prominent candidate in our case is the saturated phase of the MRI, but further investigations are required to support this hypothesis.

The mass inflow rates might be affected by various other parameters. For instance, the initial mass ratio $q\, {=}\, M_2/M_1$ of the binary can affect the magnetically driven outflows. In our scenario involving the NS companion, less material remains bound near the core binary at the onset of the magnetically propelled outflow (Fig.~\ref{fig:dyn_evolution}) and, consequently, less mass can be ejected through the magnetized outflows. Hence, this second envelope-loss channel is expected to be less important compared to the CE event with the BH. 
Likewise, the initial mass of the primary star (i.e., the amount of mass that is participating in the CE event) and its evolutionary stage (i.e., the binding energy of the envelopes, which are expected to be lower for less evolved stars) should directly impact the evolution. This might explain why \citet{ondratschek2022a} find the magnetically driven outflows to be unimportant for the overall CE ejection process. 

The lower resolution in low-density regions because of the mass-adapted unstructured mesh generated by \textsc{arepo} implies rather poor resolution of the pressure scale height (we find $3\, \textit{--} \, 60$ cells per $H$) in the CBD material (Sect.~\ref{sec:CBD}). This can enhance numerical viscosity and may lead to overestimated $\alpha$-viscosities.

%%%%%%%%%%%%%%%%%%%%%%%%%%%%%%%%%%%%%%%%%%%%%%%%%%%%%%%%%%%%%%%%%%%%%
\section{Conclusions}\label{sec:conclusion}
In this paper, we turned our attention once more to the CE interaction of a $10\,\msol$ RSG with a $5\, \msol$ BH as well as a $1.4\, \msol$ NS companion, that was first simulated by \citet{moreno2022a}. Utilizing a new refinement approach for the softened primary core and companion particles, we were able to follow the dynamic evolution until full kinetic envelope ejection. Our main results can be summarized as follows:
\begin{itemize}
    \item After 1220 (1831) orbits, a final orbital separation of $a_\mathrm{f,\, BH} \,{\approx}\, 37 \, \rsol$ ($a_\mathrm{f, \, NS} \,{\approx}\, 14 \, \rsol$) is established for the BH (NS) companion. In terms of the energy prescription, these final distances translate to $\alpha_\mathrm{CE, \, BH} \,{=}\, 0.50$ and $\alpha_\mathrm{CE, \, NS} \,{=}\, 0.57$ for the BH and NS companions, respectively. 
    Despite full envelope ejection, the orbital distances we obtain in our simulations exceed the expected separations of $a_\mathrm{f, \, \mathrm{BH}} \,{\approx}\, 19.2 \, \rsol$ and $a_\mathrm{f\, \mathrm{NS}} \,{\approx}\, 5.9 \, \rsol$ by the $\alpha$-formalism (here $\alpha_\mathrm{th}\,{=}\,1, \, \alpha_\mathrm{CE}\,{=}\,1$). In agreement with simulations of CE interactions in low-mass systems \citep[e.g.,][]{sand2020a}, the envelope ejection induced by recombination in the post-spiral-in systems has little effect on the final orbital separations.
    \item The morphology of our system changes significantly during our conducted CE simulations. The initial perturbation of the envelope during the spiral-in of the companions breaks the spherical symmetry of the primary star and leaves behind a system imprinted by the symmetry of the orbital motion of the core binary. Consequently, after the spiral-in, the bound material forms a toroidal structure visually resembling a geometrically thick CBD and fulfills the dynamic requirements (i.e., gravitationally bound, circularized motion, Raleigh stability and vertical HSE) associated with stable disk structures. The bound material is found to be eroded by two mass loss channels: radial inward mass fluxes and recombination-driven winds, which drive the ejection of envelope material in the opposite direction.    
    \item The first channel is refueling magnetically driven polar outflows appearing in both of our simulations\footnote{For a detailed discussion, we refer to Vetter et al.~in prep.} similar to those observed in simulations of lower mass systems \citep{ondratschek2022a}. They accelerate high-entropy material to peak velocities of $~300\, \mathrm{km \, s^{-1}}$, which flows along low-density channels perpendicular to the orbital plane. When the polar outflows interact with the surrounding material, bow shocks are observed. Together with the toroidal structure, they do resemble the morphological appearance of bipolar pre-PN and the magnetically driven outflow can act like a pressure valve in our systems. 
    \item From the radial (averaged) distributions of the bound material, we can infer the time evolution of physical properties such as pressure scale height, aspect ratio, temperature, column density and angular velocity. They show the characteristics expected from pressure-supported, sub-Keplerian thick disks with an aspect ratio of ${\sim}\, 0.6\,\textit{--}\, 1.0$ (see Sect.~\ref{sec:CBD}, especially Fig.~\ref{fig:disk_average_quants}). 
    We additionally estimate the radial mass fluxes through both mass loss channels (see Fig.~\ref{fig:M_dot}). 
    The inward-directed mass transport rates can be reproduced by a turbulent transport model with an $\alpha$-viscosity of $0.06\,\textit{--}\, 0.13$. It appears likely that the stress-related transport is induced by the nonlinear regime of the MRI, and most of the material is expected to be launched by the magnetically driven outflow (Vetter et al.~in prep.). 
    Further, we show that the second channel of radial winds eroding the disk-like structure in radial direction in the outer part of the CBD is driven by the local thermalization of hydrogen recombination energy. This channel is found to exceed the inward flows significantly and is thus be expected to be the dominant mass ejection channel. 
    It can potentially be affected by cooling processes, and further investigations, including cooling models, are required to refine our understanding of the properties of a post-CE CBD.
    \item We analyzed the possible final fates of the systems, similar to the analysis in \citet{moreno2022a}. In addition to the default scenario of a subsequent MT and a natal NS kick, we also accounted for the interaction with a CBD. 
    The fractions $f_\mathrm{merge}$ of mergers due to GW emission amount to $0.9$ -- $10.7 \, \%$ for the BH and NS systems when accounting for a stable MT after the CE interaction and a natal NS kick (see Sect.~\ref{sec:final_fate}). 
    The additional interaction of the engulfed binary with a CBD (Sect.~\ref{sec:final_fate}, \citealp{tuna2023a,wei2023a}), however, increases the merger fractions significantly, and we found merger fractions of approximately $10 \, (40) \, \mathrm{\%}$ for the system involving a BH (NS) companion. This agrees with the results obtained in \citet{wei2023a}, and we refer to this publication for a more detailed analysis of the possible fate of the post-CE systems resulting from our simulations.
\end{itemize}
%%%%%%%%%%%%%%%%%%%%%%%%%%%%%%%%%%%%%%%%%%%%%%%%%%%%%%%%%%%%%%%%%%%%%
\begin{acknowledgements}
    We thank the referee for the constructive and helpful comments, which helped us to improve the paper. M.V., F.K.R., F.R.N.S., M.Y.M.L., and R.A.\ acknowledge support by the Klaus-Tschira Foundation. M.Y.M.L. has been supported by a Croucher Foundation Fellowship. M.V., F.K.R., and R.A.\ acknowledge funding by the European Union (ERC, ExCEED, project number 101096243). Views and opinions expressed are, however, those of the authors only and do not necessarily reflect those of the European Union or the European Research Council Executive Agency. Neither the European Union nor the granting authority can be held responsible for them.
\end{acknowledgements} 
%-------------------------------------------------------------------
\bibliography{astrofritz_edited} 

\begin{thebibliography}{89}
\expandafter\ifx\csname natexlab\endcsname\relax\def\natexlab#1{#1}\fi

\bibitem[{Adachi {et~al.}(1976)Adachi, Hayashi, \& Nakazawa}]{adachi1976a}
Adachi, I., Hayashi, C., \& Nakazawa, K. 1976, Progress of Theoretical Physics,
  56, 1756

\bibitem[{Balbus \& Hawley(1998)}]{balbus1998a}
Balbus, S.~A. \& Hawley, J.~F. 1998, Rev. Mod. Phys., 70, 1

\bibitem[{{Belczynski} {et~al.}(2016){Belczynski}, {Holz}, {Bulik}, \&
  {O'Shaughnessy}}]{belczynski2020a}
{Belczynski}, K., {Holz}, D.~E., {Bulik}, T., \& {O'Shaughnessy}, R. 2016,
  \nat, 534, 512

\bibitem[{{Brandt} \& {Podsiadlowski}(1995)}]{brandt1995a}
{Brandt}, N. \& {Podsiadlowski}, P. 1995, \mnras, 274, 461

\bibitem[{Bronner {et~al.}(2023)Bronner, Schneider, Podsiadlowski, \&
  Röpke}]{bronner2023a}
Bronner, V., Schneider, F., Podsiadlowski, P., \& Röpke, F. 2023, \aap, 683

\bibitem[{{Chamandy} {et~al.}(2019){Chamandy}, {Blackman}, {Frank},
  {Carroll-Nellenback}, {Zou}, \& {Tu}}]{chamandy2019a}
{Chamandy}, L., {Blackman}, E.~G., {Frank}, A., {et~al.} 2019, \mnras, 490,
  3727

\bibitem[{{Chamandy} {et~al.}(2024){Chamandy}, {Carroll-Nellenback},
  {Blackman}, {Frank}, {Tu}, {Liu}, {Zou}, \& {Nordhaus}}]{chamandy2023a}
{Chamandy}, L., {Carroll-Nellenback}, J., {Blackman}, E.~G., {et~al.} 2024,
  \mnras, 528, 234

\bibitem[{{Chamandy} {et~al.}(2018){Chamandy}, {Frank}, {Blackman},
  {Carroll-Nellenback}, {Liu}, {Tu}, {Nordhaus}, {Chen}, \&
  {Peng}}]{chamandy2018a}
{Chamandy}, L., {Frank}, A., {Blackman}, E.~G., {et~al.} 2018, \mnras, 480,
  1898

\bibitem[{{Chandrasekhar}(1943)}]{chandrasekhar1943a}
{Chandrasekhar}, S. 1943, \apj, 97, 255

\bibitem[{{Courant} {et~al.}(1928){Courant}, {Friedrichs}, \&
  {Lewy}}]{courant1928a}
{Courant}, R., {Friedrichs}, K.~O., \& {Lewy}, H. 1928, Math. Ann., 100, 32

\bibitem[{Darwin(1879)}]{darwin1879a}
Darwin, G.~H. 1879, Proceedings of the Royal Society of London, 29, 168

\bibitem[{{Davis} {et~al.}(2010){Davis}, {Stone}, \& {Pessah}}]{davis2010a}
{Davis}, S.~W., {Stone}, J.~M., \& {Pessah}, M.~E. 2010, \apj, 713, 52

\bibitem[{{de Kool}(1990)}]{dekool1990a}
{de Kool}, M. 1990, \apj, 358, 189

\bibitem[{{Dewi} \& {Tauris}(2000)}]{dewi2000a}
{Dewi}, J.~D.~M. \& {Tauris}, T.~M. 2000, \aap, 360, 1043

\bibitem[{{Dittmann} \& {Ryan}(2022)}]{dittmann2022a}
{Dittmann}, A.~J. \& {Ryan}, G. 2022, \mnras, 513, 6158

\bibitem[{{Dokuchaev}(1964)}]{dokuchaev1964a}
{Dokuchaev}, V.~P. 1964, \sovast, 8, 23

\bibitem[{{Dominik} {et~al.}(2012){Dominik}, {Belczynski}, {Fryer}, {Holz},
  {Berti}, {Bulik}, {Mandel}, \& {O'Shaughnessy}}]{dominik2012a}
{Dominik}, M., {Belczynski}, K., {Fryer}, C., {et~al.} 2012, \apj, 759, 52

\bibitem[{Eggleton(2011)}]{eggleton2011a}
Eggleton, P. 2011, Evolutionary Processes in Binary and Multiple Stars
  (Cambridge: Cambridge University Press)

\bibitem[{Gagnier \& Pejcha(2023)}]{gagnier2023a}
Gagnier, D. \& Pejcha, O. 2023, A\&A, 674, A121

\bibitem[{Gagnier \& Pejcha(2024)}]{gagnier2024a}
Gagnier, D. \& Pejcha, O. 2024, A\&A, 683, A4

\bibitem[{Genel {et~al.}(2013)Genel, Vogelsberger, Nelson, Sijacki, Springel,
  \& Hernquist}]{genel2013a}
Genel, S., Vogelsberger, M., Nelson, D., {et~al.} 2013, Monthly Notices of the
  Royal Astronomical Society, 435, 1426

\bibitem[{{Giacobbo} \& {Mapelli}(2018)}]{giacobbo2018a}
{Giacobbo}, N. \& {Mapelli}, M. 2018, \mnras, 480, 2011

\bibitem[{{Hirai} \& {Mandel}(2022)}]{hirai2022a}
{Hirai}, R. \& {Mandel}, I. 2022, \apjl, 937, L42

\bibitem[{{Hobbs} {et~al.}(2005){Hobbs}, {Lorimer}, {Lyne}, \&
  {Kramer}}]{hobbs2005a}
{Hobbs}, G., {Lorimer}, D.~R., {Lyne}, A.~G., \& {Kramer}, M. 2005, \mnras,
  360, 974

\bibitem[{{Hurley} {et~al.}(2002){Hurley}, {Tout}, \& {Pols}}]{hurley2002a}
{Hurley}, J.~R., {Tout}, C.~A., \& {Pols}, O.~R. 2002, \mnras, 329, 897

\bibitem[{{Iaconi} {et~al.}(2017){Iaconi}, {Reichardt}, {Staff}, {De Marco},
  {Passy}, {Price}, {Wurster}, \& Herwig}]{iaconi2017a}
{Iaconi}, R., {Reichardt}, T., {Staff}, J., {et~al.} 2017, \mnras, 464, 4028

\bibitem[{{Iglesias} \& {Rogers}(1996)}]{iglesias1996a}
{Iglesias}, C.~A. \& {Rogers}, F.~J. 1996, \apj, 464, 943

\bibitem[{{Ivanova}(2011)}]{ivanova2011a}
{Ivanova}, N. 2011, \apj, 730, 76

\bibitem[{{Ivanova} {et~al.}(2013){Ivanova}, {Justham}, {Chen}, {De Marco},
  {Fryer}, {Gaburov}, {Ge}, {Glebbeek}, {Han}, {Li}, {Lu}, {Marsh},
  {Podsiadlowski}, {Potter}, {Soker}, {Taam}, {Tauris}, {van den Heuvel}, \&
  {Webbink}}]{ivanova2013a}
{Ivanova}, N., {Justham}, S., {Chen}, X., {et~al.} 2013, \aapr, 21, 59

\bibitem[{Jiang {et~al.}(2021)Jiang, Tauris, Chen, \& Fuller}]{jiang2021a}
Jiang, L., Tauris, T.~M., Chen, W.-C., \& Fuller, J. 2021, The Astrophysical
  Journal Letters, 920, L36

\bibitem[{{Kim} \& {Kim}(2007)}]{kim2007a}
{Kim}, H. \& {Kim}, W.-T. 2007, \apj, 665, 432

\bibitem[{{Kim}(2010)}]{kim2010a}
{Kim}, W.-T. 2010, \apj, 725, 1069

\bibitem[{{Kramer} {et~al.}(2020){Kramer}, {Schneider}, {Ohlmann}, {Geier},
  {Schaffenroth}, {Pakmor}, \& {R{\"o}pke}}]{kramer2020a}
{Kramer}, M., {Schneider}, F.~R.~N., {Ohlmann}, S.~T., {et~al.} 2020, \aap,
  642, A97

\bibitem[{{Kruckow} {et~al.}(2018){Kruckow}, {Tauris}, {Langer}, {Kramer}, \&
  {Izzard}}]{kruckow2018a}
{Kruckow}, M.~U., {Tauris}, T.~M., {Langer}, N., {Kramer}, M., \& {Izzard},
  R.~G. 2018, \mnras, 481, 1908

\bibitem[{{Laplace} {et~al.}(2020){Laplace}, {G{\"o}tberg}, {de Mink},
  {Justham}, \& {Farmer}}]{laplace2020b}
{Laplace}, E., {G{\"o}tberg}, Y., {de Mink}, S.~E., {Justham}, S., \& {Farmer},
  R. 2020, \aap, 637, A6

\bibitem[{{Lau} {et~al.}(2022{\natexlab{a}}){Lau}, {Hirai},
  {Gonz{\'a}lez-Bol{\'\i}var}, {Price}, {De Marco}, \& {Mandel}}]{lau2022a}
{Lau}, M. Y.~M., {Hirai}, R., {Gonz{\'a}lez-Bol{\'\i}var}, M., {et~al.}
  2022{\natexlab{a}}, \mnras, 512, 5462

\bibitem[{{Lau} {et~al.}(2022{\natexlab{b}}){Lau}, {Hirai}, {Price}, \&
  {Mandel}}]{lau2022b}
{Lau}, M. Y.~M., {Hirai}, R., {Price}, D.~J., \& {Mandel}, I.
  2022{\natexlab{b}}, \mnras, 516, 4669

\bibitem[{{Law-Smith} {et~al.}(2020){Law-Smith}, {Everson}, {Ramirez-Ruiz}, {de
  Mink}, {van Son}, {G{\"o}tberg}, {Zellmann}, {Vigna-G{\'o}mez}, {Renzo},
  {Wu}, {Schr{\o}der}, {Foley}, \& {Hutchinson-Smith}}]{law-smith2020a}
{Law-Smith}, J. A.~P., {Everson}, R.~W., {Ramirez-Ruiz}, E., {et~al.} 2020,
  arXiv e-prints, arXiv:2011.06630

\bibitem[{Mandel \& Broekgaarden(2022)}]{mandel2021a}
Mandel, I. \& Broekgaarden, F.~S. 2022, Living Reviews in Relativity, 25

\bibitem[{{Mapelli} \& {Giacobbo}(2018)}]{mapelli2018a}
{Mapelli}, M. \& {Giacobbo}, N. 2018, \mnras, 479, 4391

\bibitem[{{Moreno} {et~al.}(2022){Moreno}, {Schneider}, {R{\"o}pke}, {Ohlmann},
  {Pakmor}, {Podsiadlowski}, \& {Sand}}]{moreno2022a}
{Moreno}, M.~M., {Schneider}, F. R.~N., {R{\"o}pke}, F.~K., {et~al.} 2022,
  \aap, 667, A72

\bibitem[{{Mu{\~n}oz} {et~al.}(2020){Mu{\~n}oz}, {Lai}, {Kratter}, \&
  {Miranda}}]{munoz2020a}
{Mu{\~n}oz}, D.~J., {Lai}, D., {Kratter}, K., \& {Miranda}, R. 2020, \apj, 889,
  114

\bibitem[{{Narayan} \& {Yi}(1995)}]{narayan1995a}
{Narayan}, R. \& {Yi}, I. 1995, \apj, 444, 231

\bibitem[{Nixon \& King(2016)}]{nixon2016}
Nixon, C. \& King, A. 2016, Warp Propagation in Astrophysical Discs, ed.
  F.~Haardt, V.~Gorini, U.~Moschella, A.~Treves, \& M.~Colpi (Cham: Springer
  International Publishing), 45--63

\bibitem[{{Ohlmann} {et~al.}(2016{\natexlab{a}}){Ohlmann}, {R{\"o}pke},
  {Pakmor}, \& {Springel}}]{ohlmann2016a}
{Ohlmann}, S.~T., {R{\"o}pke}, F.~K., {Pakmor}, R., \& {Springel}, V.
  2016{\natexlab{a}}, \apjl, 816, L9

\bibitem[{{Ohlmann} {et~al.}(2017){Ohlmann}, {R\"{o}pke}, {Pakmor}, \&
  {Springel}}]{ohlmann2017a}
{Ohlmann}, S.~T., {R\"{o}pke}, F.~K., {Pakmor}, R., \& {Springel}, V. 2017,
  \aap, 599, A5

\bibitem[{{Ohlmann} {et~al.}(2016{\natexlab{b}}){Ohlmann}, {R\"opke}, {Pakmor},
  {Springel}, \& {M\"uller}}]{ohlmann2016b}
{Ohlmann}, S.~T., {R\"opke}, F.~K., {Pakmor}, R., {Springel}, V., \&
  {M\"uller}, E. 2016{\natexlab{b}}, \mnras, 462, L121

\bibitem[{{Ondratschek} {et~al.}(2022){Ondratschek}, {R{\"o}pke}, {Schneider},
  {Fendt}, {Sand}, {Ohlmann}, {Pakmor}, \& {Springel}}]{ondratschek2022a}
{Ondratschek}, P.~A., {R{\"o}pke}, F.~K., {Schneider}, F. R.~N., {et~al.} 2022,
  \aap, 660, L8

\bibitem[{{Ostriker}(1999)}]{ostriker1999a}
{Ostriker}, E.~C. 1999, \apj, 513, 252

\bibitem[{{Pakmor} {et~al.}(2011){Pakmor}, {Bauer}, \&
  {Springel}}]{pakmor2011d}
{Pakmor}, R., {Bauer}, A., \& {Springel}, V. 2011, \mnras, 418, 1392

\bibitem[{{Pakmor} \& {Springel}(2013)}]{pakmor2013b}
{Pakmor}, R. \& {Springel}, V. 2013, \mnras, 432, 176

\bibitem[{{Passy} {et~al.}(2012){Passy}, {De Marco}, {Fryer}, {Herwig},
  {Diehl}, {Oishi}, {Mac Low}, {Bryan}, \& {Rockefeller}}]{passy2012a}
{Passy}, J.-C., {De Marco}, O., {Fryer}, C.~L., {et~al.} 2012, \apj, 744, 52

\bibitem[{{Paxton} {et~al.}(2011){Paxton}, {Bildsten}, {Dotter}, {Herwig},
  {Lesaffre}, \& {Timmes}}]{paxton2011a}
{Paxton}, B., {Bildsten}, L., {Dotter}, A., {et~al.} 2011, \apjs, 192, 3

\bibitem[{{Paxton} {et~al.}(2013){Paxton}, {Cantiello}, {Arras}, {Bildsten},
  {Brown}, {Dotter}, {Mankovich}, {Montgomery}, {Stello}, {Timmes}, \&
  {Townsend}}]{paxton2013a}
{Paxton}, B., {Cantiello}, M., {Arras}, P., {et~al.} 2013, \apjs, 208, 4

\bibitem[{{Paxton} {et~al.}(2015){Paxton}, {Marchant}, {Schwab}, {Bauer},
  {Bildsten}, {Cantiello}, {Dessart}, {Farmer}, {Hu}, {Langer}, {Townsend},
  {Townsley}, \& {Timmes}}]{paxton2015a}
{Paxton}, B., {Marchant}, P., {Schwab}, J., {et~al.} 2015, \apjs, 220, 15

\bibitem[{{Paxton} {et~al.}(2018){Paxton}, {Schwab}, {Bauer}, {Bildsten},
  {Blinnikov}, {Duffell}, {Farmer}, {Goldberg}, {Marchant}, {Sorokina},
  {Thoul}, {Townsend}, \& {Timmes}}]{paxton2018a}
{Paxton}, B., {Schwab}, J., {Bauer}, E.~B., {et~al.} 2018, \apjs, 234, 34

\bibitem[{{Paxton} {et~al.}(2019){Paxton}, {Smolec}, {Schwab}, {Gautschy},
  {Bildsten}, {Cantiello}, {Dotter}, {Farmer}, {Goldberg}, {Jermyn}, {Kanbur},
  {Marchant}, {Thoul}, {Townsend}, {Wolf}, {Zhang}, \& {Timmes}}]{paxton2019a}
{Paxton}, B., {Smolec}, R., {Schwab}, J., {et~al.} 2019, \apjs, 243, 10

\bibitem[{Peters(1964)}]{peters1964a}
Peters, P.~C. 1964, Phys. Rev., 136, B1224

\bibitem[{{Planck Collaboration} {et~al.}(2020){Planck Collaboration},
  {Aghanim}, {Akrami}, {Ashdown}, {Aumont}, {Baccigalupi}, {Ballardini},
  {Banday}, {Barreiro}, {Bartolo}, {Basak}, {Battye}, {Benabed}, {Bernard},
  {Bersanelli}, {Bielewicz}, {Bock}, {Bond}, {Borrill}, {Bouchet}, {Boulanger},
  {Bucher}, {Burigana}, {Butler}, {Calabrese}, {Cardoso}, {Carron},
  {Challinor}, {Chiang}, {Chluba}, {Colombo}, {Combet}, {Contreras}, {Crill},
  {Cuttaia}, {de Bernardis}, {de Zotti}, {Delabrouille}, {Delouis}, {Di
  Valentino}, {Diego}, {Dor{\'e}}, {Douspis}, {Ducout}, {Dupac}, {Dusini},
  {Efstathiou}, {Elsner}, {En{\ss}lin}, {Eriksen}, {Fantaye}, {Farhang},
  {Fergusson}, {Fernandez-Cobos}, {Finelli}, {Forastieri}, {Frailis},
  {Fraisse}, {Franceschi}, {Frolov}, {Galeotta}, {Galli}, {Ganga},
  {G{\'e}nova-Santos}, {Gerbino}, {Ghosh}, {Gonz{\'a}lez-Nuevo}, {G{\'o}rski},
  {Gratton}, {Gruppuso}, {Gudmundsson}, {Hamann}, {Handley}, {Hansen},
  {Herranz}, {Hildebrandt}, {Hivon}, {Huang}, {Jaffe}, {Jones}, {Karakci},
  {Keih{\"a}nen}, {Keskitalo}, {Kiiveri}, {Kim}, {Kisner}, {Knox},
  {Krachmalnicoff}, {Kunz}, {Kurki-Suonio}, {Lagache}, {Lamarre}, {Lasenby},
  {Lattanzi}, {Lawrence}, {Le Jeune}, {Lemos}, {Lesgourgues}, {Levrier},
  {Lewis}, {Liguori}, {Lilje}, {Lilley}, {Lindholm}, {L{\'o}pez-Caniego},
  {Lubin}, {Ma}, {Mac{\'\i}as-P{\'e}rez}, {Maggio}, {Maino}, {Mandolesi},
  {Mangilli}, {Marcos-Caballero}, {Maris}, {Martin}, {Martinelli},
  {Mart{\'\i}nez-Gonz{\'a}lez}, {Matarrese}, {Mauri}, {McEwen}, {Meinhold},
  {Melchiorri}, {Mennella}, {Migliaccio}, {Millea}, {Mitra},
  {Miville-Desch{\^e}nes}, {Molinari}, {Montier}, {Morgante}, {Moss}, {Natoli},
  {N{\o}rgaard-Nielsen}, {Pagano}, {Paoletti}, {Partridge}, {Patanchon},
  {Peiris}, {Perrotta}, {Pettorino}, {Piacentini}, {Polastri}, {Polenta},
  {Puget}, {Rachen}, {Reinecke}, {Remazeilles}, {Renzi}, {Rocha}, {Rosset},
  {Roudier}, {Rubi{\~n}o-Mart{\'\i}n}, {Ruiz-Granados}, {Salvati}, {Sandri},
  {Savelainen}, {Scott}, {Shellard}, {Sirignano}, {Sirri}, {Spencer},
  {Sunyaev}, {Suur-Uski}, {Tauber}, {Tavagnacco}, {Tenti}, {Toffolatti},
  {Tomasi}, {Trombetti}, {Valenziano}, {Valiviita}, {Van Tent}, {Vibert},
  {Vielva}, {Villa}, {Vittorio}, {Wandelt}, {Wehus}, {White}, {White},
  {Zacchei}, \& {Zonca}}]{planckcollab2020a}
{Planck Collaboration}, {Aghanim}, N., {Akrami}, Y., {et~al.} 2020, \aap, 641,
  A6

\bibitem[{{Podsiadlowski} {et~al.}(2004){Podsiadlowski}, {Langer},
  {Poelarends}, {Rappaport}, {Heger}, \& {Pfahl}}]{podsiadlowski2004a}
{Podsiadlowski}, P., {Langer}, N., {Poelarends}, A.~J.~T., {et~al.} 2004, \apj,
  612, 1044

\bibitem[{{Podsiadlowski} {et~al.}(2003){Podsiadlowski}, {Rappaport}, \&
  {Han}}]{podsiadlowski2003a}
{Podsiadlowski}, P., {Rappaport}, S., \& {Han}, Z. 2003, \mnras, 341, 385

\bibitem[{{Podsiadlowski} {et~al.}(2002){Podsiadlowski}, {Rappaport}, \&
  {Pfahl}}]{podsiadlowski2002a}
{Podsiadlowski}, P., {Rappaport}, S., \& {Pfahl}, E.~D. 2002, \apj, 565, 1107

\bibitem[{{Powell} {et~al.}(1999){Powell}, {Roe}, {Linde}, {Gombosi}, \& {de
  Zeeuw}}]{powell1999a}
{Powell}, K.~G., {Roe}, P.~L., {Linde}, T.~J., {Gombosi}, T.~I., \& {de Zeeuw},
  D.~L. 1999, Journal of Computational Physics, 154, 284

\bibitem[{{Pringle}(1981)}]{pringle1981a}
{Pringle}, J.~E. 1981, \araa, 19, 137

\bibitem[{Pringle \& King(2007)}]{pringle2007a}
Pringle, J.~E. \& King, A. 2007, Astrophysical Flows (Cambridge University
  Press)

\bibitem[{{Prust} \& {Chang}(2019)}]{prust2019a}
{Prust}, L.~J. \& {Chang}, P. 2019, \mnras, 486, 5809

\bibitem[{{Reichardt} {et~al.}(2020){Reichardt}, {De Marco}, {Iaconi},
  {Chamandy}, \& {Price}}]{reichardt2020a}
{Reichardt}, T.~A., {De Marco}, O., {Iaconi}, R., {Chamandy}, L., \& {Price},
  D.~J. 2020, \mnras, 494, 5333

\bibitem[{{Ricker} \& {Taam}(2012)}]{ricker2012a}
{Ricker}, P.~M. \& {Taam}, R.~E. 2012, \apj, 746, 74

\bibitem[{{Rogers} \& {Nayfonov}(2002)}]{rogers2002a}
{Rogers}, F.~J. \& {Nayfonov}, A. 2002, \apj, 576, 1064

\bibitem[{{Rogers} {et~al.}(1996){Rogers}, {Swenson}, \&
  {Iglesias}}]{rogers1996a}
{Rogers}, F.~J., {Swenson}, F.~J., \& {Iglesias}, C.~A. 1996, \apj, 456, 902

\bibitem[{{R{\"o}pke} \& {De Marco}(2023)}]{roepke2022a}
{R{\"o}pke}, F.~K. \& {De Marco}, O. 2023, Living Reviews in Computational
  Astrophysics, 9, 2

\bibitem[{{Sand} {et~al.}(2020){Sand}, {Ohlmann}, {Schneider}, {Pakmor}, \&
  {R{\"o}pke}}]{sand2020a}
{Sand}, C., {Ohlmann}, S.~T., {Schneider}, F. R.~N., {Pakmor}, R., \&
  {R{\"o}pke}, F.~K. 2020, \aap, 644, A60

\bibitem[{Schneider {et~al.}(2021)Schneider, Podsiadlowski, \&
  M{\"u}ller}]{schneider2021a}
Schneider, F., Podsiadlowski, P., \& M{\"u}ller, B. 2021, A\&A, 645

\bibitem[{{Shakura} \& {Sunyaev}(1973)}]{shakura1973a}
{Shakura}, N.~I. \& {Sunyaev}, R.~A. 1973, \aap, 24, 337

\bibitem[{{Shi} {et~al.}(2012){Shi}, {Krolik}, {Lubow}, \& {Hawley}}]{shi2012a}
{Shi}, J.-M., {Krolik}, J.~H., {Lubow}, S.~H., \& {Hawley}, J.~F. 2012, \apj,
  749, 118

\bibitem[{{Siwek} {et~al.}(2023{\natexlab{a}}){Siwek}, {Weinberger}, \&
  {Hernquist}}]{siwek2023a}
{Siwek}, M., {Weinberger}, R., \& {Hernquist}, L. 2023{\natexlab{a}}, \mnras,
  522, 2707

\bibitem[{{Siwek} {et~al.}(2023{\natexlab{b}}){Siwek}, {Weinberger},
  {Mu{\~n}oz}, \& {Hernquist}}]{siwek2022a}
{Siwek}, M., {Weinberger}, R., {Mu{\~n}oz}, D.~J., \& {Hernquist}, L.
  2023{\natexlab{b}}, \mnras, 518, 5059

\bibitem[{{Springel}(2010)}]{springel2010a}
{Springel}, V. 2010, \mnras, 401, 791

\bibitem[{Stevenson {et~al.}(2017)Stevenson, Vigna-G{\'o}mez, Mandel, Barrett,
  Neijssel, Perkins, \& de~Mink}]{stevenson2017a}
Stevenson, S., Vigna-G{\'o}mez, A., Mandel, I., {et~al.} 2017, Nature
  Communications, 8, 14906

\bibitem[{{Tauris}(1996)}]{tauris1996a}
{Tauris}, T.~M. 1996, \aap, 315, 453

\bibitem[{{Tauris} {et~al.}(2017){Tauris}, {Kramer}, {Freire}, {Wex}, {Janka},
  {Langer}, {Podsiadlowski}, {Bozzo}, {Chaty}, {Kruckow}, {van den Heuvel},
  {Antoniadis}, {Breton}, \& {Champion}}]{tauris2017a}
{Tauris}, T.~M., {Kramer}, M., {Freire}, P.~C.~C., {et~al.} 2017, \apj, 846,
  170

\bibitem[{Tiede {et~al.}(2020)Tiede, Zrake, MacFadyen, \& Haiman}]{tiede2020a}
Tiede, C., Zrake, J., MacFadyen, A., \& Haiman, Z. 2020, The Astrophysical
  Journal, 900, 43

\bibitem[{{Tuna} \& {Metzger}(2023)}]{tuna2023a}
{Tuna}, S. \& {Metzger}, B.~D. 2023, \apj, 955, 125

\bibitem[{{Vigna-G{\'o}mez} {et~al.}(2018){Vigna-G{\'o}mez}, {Neijssel},
  {Stevenson}, {Barrett}, {Belczynski}, {Justham}, {de Mink}, {M{\"u}ller},
  {Podsiadlowski}, {Renzo}, {Sz{\'e}csi}, \& {Mandel}}]{vigna2018a}
{Vigna-G{\'o}mez}, A., {Neijssel}, C.~J., {Stevenson}, S., {et~al.} 2018,
  \mnras, 481, 4009

\bibitem[{Vigna-Gómez {et~al.}(2022)Vigna-Gómez, Wassink, Klencki, Istrate,
  Nelemans, \& Mandel}]{vigna2022a}
Vigna-Gómez, A., Wassink, M., Klencki, J., {et~al.} 2022, Monthly Notices of
  the Royal Astronomical Society, 511, 2326

\bibitem[{{Webbink}(1984)}]{webbink1984a}
{Webbink}, R.~F. 1984, \apj, 277, 355

\bibitem[{{Wei} {et~al.}(2024){Wei}, {Schneider}, {Podsiadlowski}, {Laplace},
  {R{\"o}pke}, \& {Vetter}}]{wei2023a}
{Wei}, D., {Schneider}, F. R.~N., {Podsiadlowski}, P., {et~al.} 2024, \aap,
  688, A87

\bibitem[{{Weidenschilling}(1977)}]{weidenschilling1977a}
{Weidenschilling}, S.~J. 1977, \apss, 51, 153

\bibitem[{{Woosley}(2019)}]{woosley2019a}
{Woosley}, S.~E. 2019, \apj, 878, 49

\end{thebibliography}

\clearpage
\begin{appendix}
\twocolumn[
\begin{@twocolumnfalse}
\section*{\centering Appendices}
\end{@twocolumnfalse}
]
%--------------------------------------------------------------------
\section{A new refinement approach for gravity-only particles in CE interaction}\label{sec:appendix_refinement}
\begin{figure}
    \centering
    \resizebox{\hsize}{!} {\includegraphics{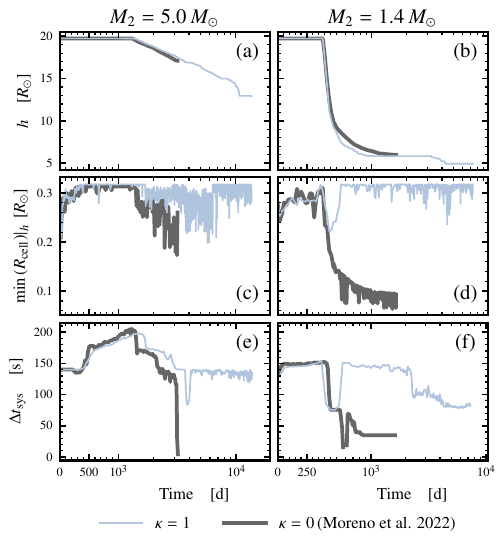}}
    \caption{Temporal evolution of the softening length $h$ in (a) and (b), the smallest cell radius within the softened spheres $\min{(R_\mathrm{cell})}|_h$ in (c) and (d) as well as the system time step $\Delta t_\mathrm{sys}$ in (e) and (f). The simulations conducted with $\kappa=1$ (see Sect.~\ref{sec:refinement}) are shown in light blue, while the simulations with $\kappa=0$ \citep{moreno2022a} are shown in black. The plots on the left feature the system involving a BH companion and the plots on the right the CE models involving the NS companion, respectively.}
    \label{fig:appendixB_comp_costs}
\end{figure}
In Sect.~\ref{sec:refinement}, we introduced a new refinement approach within the softened gravitational potential of the two point particles representing the core of the primary star and the companion. As mentioned, the softening length $h$ is reduced over the course of our CE simulations (Fig.~\ref{fig:appendixB_comp_costs}a and \ref{fig:appendixB_comp_costs}b) as the two point particles approach each other in order to avoid a potential overlap of the softened spheres. As a consequence of the implementation of this procedure, the softening length can be expressed as a monotonic decreasing function depending on the reduction given by $a(t)$ (Fig.~\ref{fig:appendixB_comp_costs}). Combined with Eq.~(\ref{eq:N_CPS}), it is immediately apparent that the cell sizes within the softened potential ($R_\mathrm{cell} \,{\approx}\, h(t)/N_\mathrm{CPS}$) must be constant for $\kappa\,{=}\,1$ and decrease in the $\kappa\,{=}\,0$ case (Fig.~\ref{fig:appendixB_comp_costs}c and \ref{fig:appendixB_comp_costs}d). 

The system time step $\Delta t_\mathrm{sys}$ in Fig.~\ref{fig:appendixB_comp_costs}e and \ref{fig:appendixB_comp_costs}f seems to reflect the behavior of the smallest cell sizes according to the CFL condition \citep{courant1928a} for the NS case. But we also observe an unexpectedly sharp decline in the system involving the BH companions for the simulations with $\kappa\,{=}\,0$ (black line at around $3000 \, \mathrm{d}$). This steep drop can be explained by the onset of the magnetized outflow (Vetter et al.~in prep.), in which strongly magnetized and high velocity but low density material (i.e., large cell sizes) propagates through a rather weakly magnetized gas. Given the implementation of magnetohydrodynamics in \textsc{arepo} (see Sect.~\ref{sec:methods}), the relatively harsh difference in magnetization between the magnetically driven outflow and ambient material provokes a steep drop in the system time step. Additionally, for the NS companion and $\kappa\,{=}\,0$ (black line), there is a small recovery of $\Delta t_\mathrm{sys}$ at around, $500\, \mathrm{d}$ which is caused by a restart of the simulation with a higher $C_\mathrm{CFL}$-factor. This adjustment increased the system time step but compared to $\kappa=0$, the new refinement approach allows us to use time steps four times longer. 

Over the course of the simulations, the number of cells per pressure scale height in the BH (NS) case decreases from ${\approx}\, 11$ to ${\approx}\, 7$ (${\approx}\, 3$) for $\kappa\,{=}\,1$ and ${\approx}\, 5$ (${\approx}\, 7$) for $\kappa\,{=}\,0$, respectively. These numbers may raise concerns and, for example, \citet{ohlmann2017a} suggests a linear resolution of the pressure scale height of $3.2\,\textit{--}\, 6.8$ cells for a stabilized atmosphere down to mach numbers of $\mathcal{M} \,{=}\, 0.01\, \textit{--} \, 0.001$ (with adiabatic index $\gamma \,{=}\, 5/3$ and CFL factor of $C_\mathrm{CFL} \,{=}\, 0.4$). But this neither refers to the stellar remnant in our model, nor does it take the modified structure of the core into account. 
In conclusion, the new refinement approach mitigates the issue of decreasing time steps in the late stages of the modeled CE interactions, but the constant cell sizes may also introduce additional challenges in achieving convergence in orbital separation and envelope ejection with respect to the choice of the initial cut radius.

%--------------------------------------------------------------------
\section{Influence of the new refinement criterion on the orbital separation}\label{sec:appendix_numerical_uncertanties}
\begin{figure}
    \centering
    \resizebox{\hsize}{!} {\includegraphics{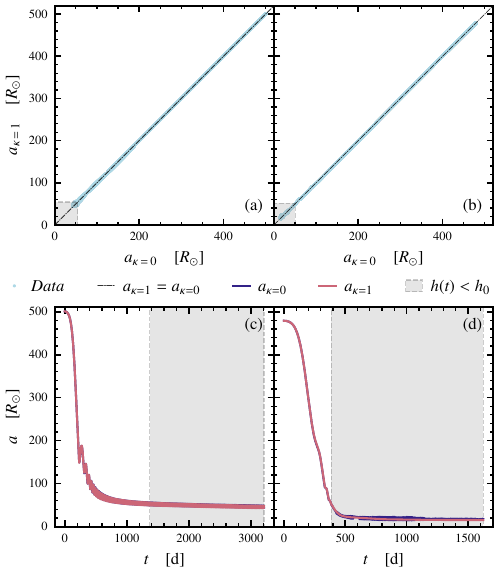}}
    \caption{Comparison between the orbital separations in our simulation with $\kappa = 1$ and the fiducial model with $\kappa = 0$ \citep{moreno2022a}, for the simulation involving the BH (left panel) and NS (right panel) companion, respectively. In (a) and (b) a scatter plot of the orbital separations ($a_\mathrm{\kappa\,{=}\,1}$ and $a_\mathrm{\kappa\,{=}\,0}$) is shown. The black dashed lines represent the perfect correlation between the separations (i.e., $a_\mathrm{\kappa\,{=}\,1} \,{=}\, a_\mathrm{\kappa\,{=}\,0}$), while the best linear fit is $a_\mathrm{\kappa\,{=}\,1} \,{\approx}\, 0.998{\times} a_\mathrm{\kappa\,{=}\,0} \, (\approx0.995{\times} a_\mathrm{\kappa\,{=}\,0})$ for the BH (NS) companion.
    In the bottom row (c and d), we show the zoomed in temporal evolution of the orbital separations for $\kappa\, {=} \, 0$ (red) and $\kappa\, {=} \, 1$ (blue) similar to Fig.~\ref{fig:dyn_evolution}.}
    \label{fig:compare_sims_bland_altman}
\end{figure}

\begin{figure}
    \centering
    \resizebox{\hsize}{!} {\includegraphics{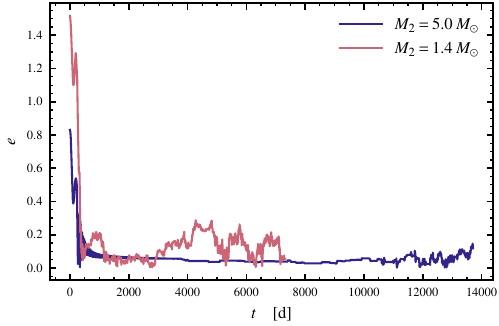}}
    \caption{Temporal evolution of the eccentricity. The simulation involving the BH companion is colored blue, while the system with the NS companion is plotted red. The eccentricity is computed via the Runge--Lenz vector.}
    \label{fig:appendixB_eccentricity}
\end{figure}

The new refinement criterion described in Sect.~\ref{sec:refinement} enabled us to follow the entire envelope ejection in both analyzed systems. When comparing our simulations to the findings of \citet{moreno2022a} (see Fig.~\ref{fig:dyn_evolution} and bottom row of  Fig.~\ref{fig:compare_sims_bland_altman}), the simulations seem to be consistent with the previous results, particularly in the earlier stages of the CE interaction. 
This affirmation gains further support when we plot the two orbital separations for $\kappa \,{=}\, 1$ and $\kappa \,{=}\, 0$ against each other (see top panels of Fig.~\ref{fig:compare_sims_bland_altman}). The best linear fit yields a slope of $\partial a_\mathrm{\kappa\,{=}\,1}/ \partial a_\mathrm{\kappa\,{=}\,0} \,{\approx}\, 0.998\, ({\approx}\, 0.995)$ for the simulations involving a BH (NS) companion.
However, after the rapid spiral-in phase in both scenarios, the orbital separations begin to exhibit artificial eccentricities, resulting in deviations from our reference simulations. In total, the (averaged) orbital separations at the time when \citet{moreno2022a} terminated their simulations are $a_{\kappa = 1\,  \mathrm{BH}}\, {=} 45.6 \, R_\odot$ and $a_{\kappa = 0\,  \mathrm{BH}}\, {=} 46.2 \, R_\odot$ for the BH companion and $a_{\kappa = 1\,  \mathrm{BH}}\, {=} 16.9 \, R_\odot$ as well as $a_{\kappa = 0\,  \mathrm{BH}}\, {=} 15.2 \, R_\odot$ for the NS companion.

The variations in eccentricity are most likely due to the sudden change of the gravitational potential, induced by the decrease of the softening length. At the point of reduction, the cells within the former softened spheres might be pushed out of HSE and (depending on the decline in $h$ and the position of the cells in the softened sphere) change their gravitational potential instantaneously, resulting in a restructuring of the density distribution in the vicinity of the point particle. In contrary to this argument, the total energy of the system is preserved down to per mil level and this effect can thus only have a minor effect on the system. Moreover, the realized cells must exhibit a significant, spatially asymmetric mass distribution at the moment of reduction to exert a kick on the binary.     
And last but potentially the strongest argument, the arising eccentricities seen in our simulations appear to be more pronounced in the late stages of the simulation. There, the temporal evolution of the softening length and the arising changes in eccentricity (e.g., Fig.~\ref{fig:dyn_evolution}) seem to show no direct correlation, and while the softening length remains constant, clear changes in eccentricity can be observed (see qualitatively in Fig.~\ref{fig:dyn_evolution}b). Regardless, the numerical approximation of the core region presents a source of uncertainty in current state-of-the-art simulations of massive CE interactions. Addressing this challenge is crucial for improving such simulations. 

%--------------------------------------------------------------------
\section{Movies of the CE interactions}\label{sec:appendix_movies}

\begin{table*}[]
    \centering
    \setlength{\tabcolsep}{8.2pt}
    \caption{Movies of the density evolution of the CE interactions involving a BH (M1 and M2) and a NS companion (M3 and M4) in the $x\,  \text{--} \, z$ and $x \, \text{--} \, y$ planes, respectively.}
    \begin{tabular}{c|cl}
        \toprule
         ID & File name & Description\\
         \midrule
         M1 & \href{https://doi.org/10.5281/zenodo.12725510}{movie\_BH\_xz.mp4}& \makecell{Simulation involving BH companion in $x \, \text{--}\, z$ plane. Additionally shown at $2970 \, \mathrm{d}$ are: \\radial velocity $v_\mathrm{r, \, sph.}$ plus velocity streamlines\\specific entropy $s$\\contour of bound material (Eq.~\ref{eq:e_kin_crit})}\\
         \midrule
         M2 & \href{https://doi.org/10.5281/zenodo.12725631}{movie\_BH\_xy.mp4}& \makecell{Simulation involving BH companion in $x \, \text{--} \, y$ plane. Additionally shown at $5418 \, \mathrm{d}$ are: \\ ionization fraction of $\mathrm{H_{\RNum{2}}}$, $\mathrm{He_{\RNum{2}}}$, $\mathrm{He_{\RNum{3}}}$ plus velocity streamlines.}\\
         \midrule
         M3 & \href{https://doi.org/10.5281/zenodo.12725663}{movie\_NS\_xz.mp4}& Same as M1 but with the NS companion. Quantities are shown at $2565 \, \mathrm{d}$\\
         \midrule
         M4 & \href{https://doi.org/10.5281/zenodo.12725695}{movie\_NS\_xy.mp4}& Same as M2 but with the NS companion. Quantities are shown at $2015 \, \mathrm{d}$\\
         \bottomrule
    \end{tabular}
    \tablefoot{The primary and companion star are marked with a \qq{x} and \qq{+,} respectively, and the fraction of ejected mass $f_\mathrm{ej, \, kin}$ according to Eq.~(\ref{eq:e_kin_crit}) is shown in all movies.}
    \label{tab:movietable}
\end{table*}

In Table~\ref{tab:movietable} we summarize the movies of the density evolution for both simulations shown in this publication. 

\end{appendix}

\end{document}